\documentclass[aps,prb,twocolumn,superscriptaddress]{revtex4-2}
\usepackage{bm}                     % para criar símbolos matemáticos em negrito
\usepackage{graphicx}               % (moderno) para incluir e manipular imagens ( gráficos, fotos, diagramas) nos seus documentos.
\usepackage{graphics}               % (antigo) para manipulação de gráficos 
\usepackage{amsmath}                % para escrita matemática (align, gather, multline para equações multi-linha com melhor alinhamento)
\usepackage{amsfonts}               % Define fontes matemáticas específicas
\usepackage{amssymb}                % Fornece uma enorme coleção de símbolos matemáticos que não estão disponíveis pelos pacotes básicos.
\usepackage{makeidx}                % para criar um índice remissivo
\usepackage{wasysym}                % coleção de símbolos diversos, incluindo muitos símbolos não-matemáticos
\usepackage{bbold}                  % Fornece uma fonte de blackboard bold alternativa à fonte \mathbb do pacote amsfonts
\usepackage[makeroom]{cancel}       % para "cancelar" ou riscar termos em expressões matemáticas
\usepackage[utf8]{inputenc}         % Habilita a entrada de caracteres UTF-8. caracteres acentuados (ã, ç, ü, á), símbolos (→, ±, °)
\usepackage[normalem]{ulem}         % Fornece sublinhados e enfatização de texto mais versáteis
\usepackage{xcolor}                 % Permite usar cores
\usepackage{soul}                   % Fornece recursos avançados de formatação de texto
\usepackage{float}                  % controle preciso sobre a posição de flutuantes (como figuras e tabelas)
\usepackage{physics}                % Pacote mais completo para notação física
\usepackage{IEEEtrantools}          % Fornece ferramentas avançadas para formatar equações matemáticas
\usepackage{mathrsfs}

\usepackage[T1]{fontenc}
\usepackage[utf8]{inputenc}

\newcommand{\be}{\begin{equation}}
\newcommand{\ee}{\end{equation}}
\newcommand{\ba}   {\begin{eqnarray}}
\newcommand{\ea}   {\end{eqnarray}}

\usepackage[unicode=true,pdfusetitle,
 bookmarks=true,bookmarksnumbered=false,bookmarksopen=false,
 breaklinks=false,pdfborder={0 0 1},backref=false,colorlinks=false]
 {hyperref}
\hypersetup{
 colorlinks,linkcolor=blue,citecolor=blue,urlcolor=blue}

\definecolor{navy}{RGB}{0,0,128}
\definecolor{royalblue}{RGB}{65,105,225}
\definecolor{darkorange}{RGB}{204,85,0} % Laranja escuro intenso
\definecolor{forestgreen}{RGB}{34,139,34}

\begin{document}

\preprint{APS/123-QED}
%---------------------------------------------------------------------------------------------------
\title{Real-Space Spectral Approach to Orbital Magnetization}

\author{Kevin J. U. Vidarte}
\email{kevin.urcia@inl.int}
\affiliation{International Iberian Nanotechnology Laboratory (INL), Av. Mestre José Veiga, 4715-330 Braga, Portugal}
\author{Henrique P. Veiga}
\affiliation{Centro de Física das Universidades do Minho e Porto
Departamento de Física e Astronomia, Faculdade de Ciências,
Universidade do Porto, 4169-007 Porto, Portugal}
\affiliation{School of Physics, Engineering and Technology and York Centre for Quantum Technologies, University of York, York YO10 5DD, United Kingdom}
\author{João M. Viana Parente Lopes}
\affiliation{Centro de Física das Universidades do Minho e Porto
Departamento de Física e Astronomia, Faculdade de Ciências,
Universidade do Porto, 4169-007 Porto, Portugal}
\author{Ramon Cardias}
\affiliation{Centro Brasileiro de Pesquisas Físicas (CBPF), Rua Dr Xavier Sigaud 150, Urca, 22290-180, Rio de Janeiro-RJ, Brazil}
\author{Aires Ferreira}
\affiliation{School of Physics, Engineering and Technology and York Centre for Quantum Technologies, University of York, York YO10 5DD, United Kingdom}

\author{Tarik P. Cysne}
\affiliation{Instituto de F\'\i sica, Universidade Federal Fluminense, 24210-346 Niter\'oi RJ, Brazil} 

\author{Tatiana G. Rappoport}
\email{tatiana.rappoport@inl.int}
\affiliation{International Iberian Nanotechnology Laboratory (INL), Av. Mestre José Veiga, 4715-330 Braga, Portugal}
\affiliation{Centro Brasileiro de Pesquisas Físicas (CBPF), Rua Dr Xavier Sigaud 150, Urca, 22290-180, Rio de Janeiro-RJ, Brazil}
\affiliation{Physics Center of Minho and Porto Universities (CF-UM-UP),Campus of Gualtar, 4710-057, Braga, Portugal}

\date{\today}% It is always \today, today, but any date may be explicitly specified
%---------------------------------------------------------------------------------------------------
\begin{abstract}
We present a real-space spectral method for computing the orbital magnetization of crystals. Starting from the commutator form of the orbital magnetization operator, we formulate an energy-resolved spectral function that is amenable to exact Chebyshev polynomial expansions and yields the total magnetization upon integration up to the Fermi level. This avoids the need for computing eigenstates and ground-state projects, providing an efficient numerical framework that is applicable to very large systems even in the presence of disorder and finite temperature. Our approach is benchmarked on the Haldane model, finding results that are in excellent agreement with the modern $k$-space formulation of orbital magnetization. Leveraging this technique, we extend our study to systems with uncorrelated disorder and point defects, and further show that the bulk Chern number can be directly obtained from the magnetization spectral density. These results open a promising route to investigate orbital responses and topological transitions in real-space models of quantum materials with realistic complexity.
\end{abstract}
%---------------------------------------------------------------------------------------------------

%\begin{description}
%\end{description}

%\keywords{}%Use showkeys class option if keyword
\maketitle

% \tableofcontents
% \tatiana{\texttt{\string\Tatiana\{blabla...\}}}\newline
% \KEVIN{\texttt{\string\KEVIN\{blabla...\}}}\newline
% \RAMON{\texttt{\string\RAMON\{blabla...\}}}\newline
% \TARIK{\texttt{\string\TARIK\{blabla...\}}}\newline

% \red{\texttt{\string\red\{blabla...\}}}\newline
% \blue{\texttt{\string\blue\{blabla...\}}}\newline
% \darkorange{\texttt{\string\darkorange\{blabla...\}}}\newline
% \forestgreen{\texttt{\string\forestgreen\{blabla...\}}}\newline

Orbital magnetization has become increasingly important in the study of topological quantum materials, where quantum metric and Berry curvature play a central role in shaping magnetic and transport properties. 
It has been identified as a key ingredient underlying the anomalous Hall effect in twisted graphene structures \cite{Aaron2019, Liu2019, Tschirhart2021}, certain topological magnets \cite{Mei2024}, and Weyl semimetals \cite{Nakamura2024}. 
The orbital magnetization has also been suggested to play an important role in the emerging physics of altermagnets \cite{Ye2025, Sopheak2025, Bangar2025} and strongly correlated quantum materials \cite{liu2025c, Zhu2025, Kang2025}, among other systems \cite{Lage2025}.  
Thus, an accurate quantitative description of orbital magnetization is essential not only for interpreting magnetic measurements, but also electrodynamic responses to external fields \cite{Liu2019, Mei2024, Nakamura2024}, magneto-electric coupling \cite{Lu2025, Malashevich2010}, and—more recently—the generation of orbital currents in the context of orbitronics \cite{Bhowal2020, Cysne2022, Pezo2023, Busch2023, Liu2025, Cysne2025, Veneri2025}.

Although great strides have been made in understanding orbital magnetization of crystals from a quantum perspective, its practical computation remains challenging.  
The modern theory expresses it as a bulk quantity, formulated as a Brillouin-zone integral over the set of occupied Bloch states and involving geometric quantities derived from Berry-phase theory \cite{Xiao2005, Thonhauser2005, Shi2007}. 
This formulation has enabled closer agreement between theory and experiment in several elemental compounds \cite{Lopez2012, Hanke2016}. 
However, its implementation in first-principles calculations is technically demanding: current approaches rely on Wannier interpolation, which requires constructing maximally localized Wannier functions and, when the relevant bands are not isolated, performing a disentanglement procedure that involves selecting suitable energy windows. 
These steps, while well established, add complexity and may introduce some uncertainty.  
In addition, the ${\bf k}$-space formulation assumes translational invariance, limiting its applicability to systems with disorder, inhomogeneity, or large real-space supercells that arise naturally in realistic materials.

Real-space formulations offer an attractive alternative. 
A case in point is the theory of Bianco and Resta, which expresses the orbital magnetization in terms of a macroscopic average of a local magnetic dipolar density \cite{Bianco2013} and has been successfully applied to both insulating and metallic systems \cite{Wang2022, Marrazzo2016, Wang2023}. Real-space viewpoints also include the single k-point formulation of Ceresoli and Resta for large systems.~\cite{Ceresoli2007}
The real-space approach is naturally suited to disordered, inhomogeneous, and finite systems, and allows for a separation between bulk and surface contributions. 
However, this technique relies on constructing the ground-state one-body density matrix (and, hence, requires knowledge of the exact eigenstates), which severely limits its applicability to large systems and studies within real-space density-functional theory. 
Moreover, although a  local \textit{marker} for the orbital magnetization can be defined, only its average over a macroscopic region has a well-defined physical meaning \cite{Seleznev2023}.

In this work, we introduce a new real-space method for computing the orbital magnetization of any finite-size system based on the exact Chebyshev polynomial expansion of a spectral function \cite{Weise2006,Garcia2015,Ferreira2015,kite,fan2021linear,Pires2022,Castro2024,Canonico2024,Cardias2025} associated with the magnetization operator $\mathscr{M}_{z}$, also discussed in Ref. \cite{Canonico2024} in the context of orbital Hall effect. 
Instead of finding the one-body density matrix of the occupied subspace by use of the ground-state  projector, we construct a \textit{spectral representation of the energy-resolved orbital magnetization density}, to be defined below, which gives access to the orbital magnetization of a sample through a simple numerical integration over energy. Our approach employs open boundary conditions, which avoids ambiguities associated with the position operator, while the use of Chebyshev expansions enables accurate investigations of realistic systems \textit{without} diagonalization of the Hamiltonian matrix. 

For non-interacting models, the computation time scales linearly with system size \cite{Weise2006,kite,fan2021linear}, at fixed energy resolution \cite{Ferreira2015}, making it ideally suited for realistic computations of disordered materials and integration into first-principles workflows lacking explicit diagonalization \cite{Cardias2025}. 
In what follows, we validate the approach %on the archetypical model of topological orbital magnetism, the Haldane model on the honeycomb lattice, 
and discuss its advantages relative to existing reciprocal- and real-space formulations.

We start from the commutator form of the \textcolor{blue}{total} orbital magnetization for a finite
two-dimensional system~\cite{Bianco2013},
\begin{equation}
M_z = -\frac{ie}{2\hbar c\,A}
\sum_{\epsilon_n<E_F}
\langle \varphi_n |\, \hat{\mathbf r}\times[\hat H,\hat{\mathbf r}] \,| \varphi_n \rangle ,
\label{eq:1}
\end{equation}
where $\hat{\mathbf r}$ is the position operator, $A$ is the sample area, $H$ is the single-particle Hamiltonian,
$\epsilon_n$ and $|\varphi_n\rangle$ are the eigenvalues and eigenvectors of $H$, respectively, and
$E_F$ is the Fermi energy.  

The expectation value above may be written as $M_z=\sum_{\epsilon_n<E_F}\langle\varphi_n|\hat{\mathscr{M}}_{z}|\varphi_n\rangle$, where
$\hat{\mathscr{M}}_{z} =-{ie}(2\hbar c A)^{-1}(\hat{x}\hat H\hat{y}-\hat{y}\hat H\hat{x})$ is the orbital magnetization operator. %\textcolor{blue}{Throughout this work, Eq. (\ref{eq:1}) is used only for finite systems with open boundary conditions, where $\hat{\mathbf r}$ is well defined; we do not interpret any spatially resolved decomposition of Eq. \ref{eq:1} as a unique local magnetization density. While the local orbital magnetization operator, $\hat{\mathscr{M}}_{z}(\textbf{r})$, \textit{per se}, is gauge and origin dependent, its equilibrium expectation value over the entire sample is well defined and gauge invariant  [add reference, perhaps Resta's work DOI: 10.1088/0953-8984/22/12/123201, or some of the other references in the paper].}
Throughout this work, Eq. (\ref{eq:1}) is applied to finite systems with open boundary conditions, where $\hat{\mathbf r}$ is well defined. Note that while the local orbital magnetization operator, $\hat{\mathscr{M}}_{z}(\textbf{r})$, \textit{per se}, is gauge and origin dependent, its equilibrium expectation value over the entire sample is well defined and gauge-invariant \cite{Resta_2010}.

We wish to show that $M_z$ can be computed without the need for explicit evaluation of the ground-state projector, $\mathscr{P}_{\text{g.s.}}=\sum_{\epsilon_{n}<E_{F}}|\varphi_{n}\rangle\langle\varphi_{n}|$ \cite{Bianco2013}. One option is to directly expand Eq. (\ref{eq:1}) in Chebyshev polynomials. However, to avoid the step discontinuity of the Fermi-Dirac occupation function, we work instead with the spectral operator
$\hat{\mathscr{M}}_z\,\delta(E-H)$, following the standard Chebyshev framework
\cite{Weise2006}. The crux of our approach is the \textit{magnetization spectral density},
\begin{equation}
m_z(E)= \mathrm{Tr}[\hat{\mathscr{M}}_{z}\,\delta(E-\hat H)].
\label{eq:2}
\end{equation}
 Knowledge of this quantity allows to retrieve the orbital magnetization  via a straightforward and quick integration according to $M_z(E_F)=\int^{E_F} m_z(E)\,dE$.  To efficiently evaluate $m_z(E)$, we expand $\delta(E-H)$ in Chebyshev polynomials after
rescaling $\hat H\!\to\!\tilde{H}$ and $E\!\to\!\tilde{E}\in[-1,1]$.  The expansion
coefficients are the moments
$\mu_n=\mathrm{Tr}[\hat{\mathscr{M}}_{z} T_n(\tilde{H})]$, computed iteratively via the
  Chebyshev recursion procedure.  To further speed up the Chebyshev scheme, the traces are computed
stochastically according to
$\mu_n\!\approx\!R^{-1}\sum_{r=1}^R\langle r|\hat{\mathscr{M}}_{z} T_n(\tilde{H})|r\rangle$, where $|r\rangle$ is the $r$-th realization of a random vector  \cite{Weise2006}.  The resulting expression for $m_z(E)$ scales linearly with the
 number of random vectors $R$, number of lattice sites, $D$, and the number of Chebyshev iterations, $M$. This formulation is fully equivalent to evaluating the zero-temperature expectation value equal to $\mathrm{Tr}[\hat{\mathscr{M}}_{z}\mathscr{P}_{\text{g.s.}}]$, but bypasses the need for eigenstates or projection
operators.  It is therefore well suited for large-scale calculations of disordered systems and first-principles parametrized  models.
Furthermore, the spectral representation provides direct access to the
magnetization spectral density, enabling analysis as a
function of chemical potential and temperature \cite{Weise2006}.

\begin{figure}[]
\centering
\includegraphics[width=1.0\linewidth]{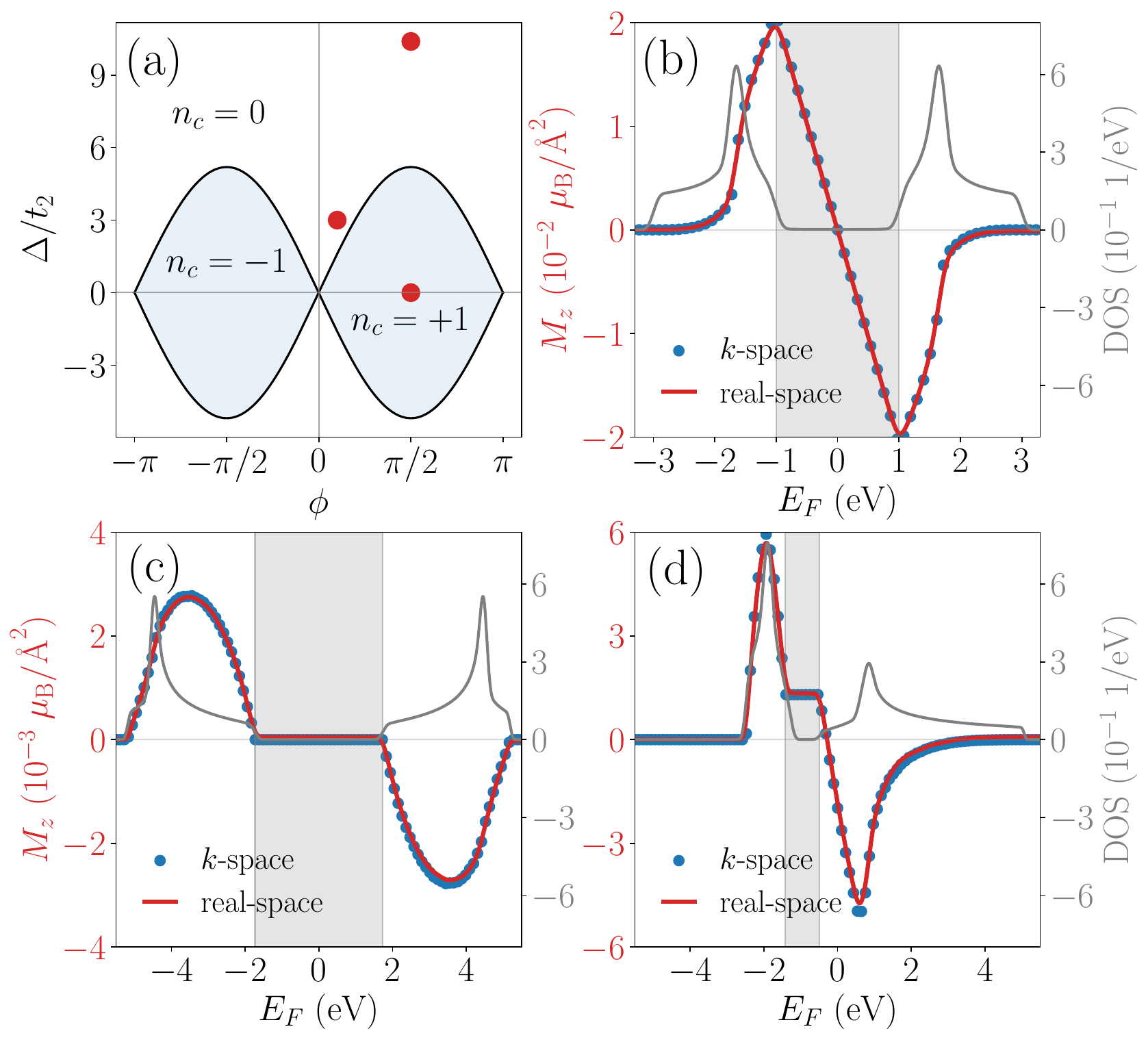}
\caption{
(a) Topological phase diagram showing the Chern number of the bottom band as a function of $\phi$ and $\Delta/t_{2}$ ($t_{1}=1$~eV and $t_{2}=t_{1}/3$).
Red points $(\phi,\Delta/t_{2})=(0.5\pi,0)$, $(0.5\pi,6\sqrt{3})$ and $(0.1\pi,3)$ indicate the parameters for panels (b), (c) and (d), respectively.
(b-d) Orbital magnetization (red lines) in real-space formulation and density of states (gray lines) as functions of the energy.
The blue dots show the results of the reciprocal-space formulation.
%The insets: Derivation of the orbital magnetization with respect to the Fermi energy inside the gap.
In our Chebyshev approach, we simulate a large rectangular domain with $10^{6}$ sites. Other parameters: $M=200$ and $R=800$.
%In our Chebyshev approach, we used a real-space system composed of $9\times 10^{4}$ sites with 200 Chebyshev polynomials and 1000 random vectors, where we use a rectangular geometry.
}
\label{fig:Fig_01}
\end{figure}

To benchmark the method we use the Haldane model on the honeycomb lattice,
a paradigmatic system for testing formulations of orbital magnetization
\cite{Haldane1988,Bianco2013}. 
It is particularly convenient because tuning the sublattice imbalance
$\Delta$, the complex next–nearest-neighbor (NNN) hopping $t_2$, and its
phase $\phi$ drives the system across topological phase transitions. 
For $\phi=\pi/2$ the model is topologically nontrivial when $|\Delta|<|3\sqrt{3}t_2|$ and
trivial otherwise; at fixed $\Delta=0$ the topology can also be switched by
varying $\phi$. 
In all regimes the system hosts a finite orbital magnetization, allowing a
direct comparison between real- and $k$-space formulations and providing an
ideal controlled setting where both trivial and topological contributions are
captured.

The real-space lattice Hamiltonian reads
\begin{equation}
\hat H = -t_1\!\!\sum_{\langle i,j\rangle} c_i^\dagger c_j
+ t_2 \!\sum_{\langle\langle i,j\rangle\rangle} e^{i\nu_{ij}\phi}\, c_i^\dagger c_j
+ \Delta \sum_i \xi_i\, c_i^\dagger c_i ,
\label{eq:haldane_real}
\end{equation}
where $t_1$ is the nearest-neighbor hopping,
$t_2 e^{i\phi}$ the complex NNN hopping with chirality
$\nu_{ij}=\pm 1$, and $\xi_i=\pm 1$ indicates the sublattice.
For $\phi=\pi/2$ the Dirac masses are
$m_{K}=\Delta-3\sqrt{3}t_2$ and $m_{K'}=\Delta+3\sqrt{3}t_2$, so the bulk
gap closes at $\Delta=\pm 3\sqrt{3}t_2$, consistent with the phase diagram
shown in Fig.~\ref{fig:Fig_01}(a).

Figures~\ref{fig:Fig_01}(b–d) compare $M_z(E_F)$ (red) obtained from the spectral method on large rectangular flakes with open boundary conditions ($708 \times 708$ unit cells, resulting in approximately $10^6$ sites) with that from the modern theory expression in $\mathbf{k}$-space (blue). The density of states is also shown (gray) to help correlate  orbital magnetization and electronic structure of the model. 
The agreement is excellent across topological and trivial insulating regimes.  
In the topological case where $\phi=\pi/2$ and $\Delta/t_2=0$ [Fig.~\ref{fig:Fig_01}(b)], the method reproduces the exactly linear dependence of $M_z$ on $E_F$ inside the bulk gap, a signature of the spectral flow of chiral edge states. 

The in-gap linear behavior, $M(E_F)\propto E_F$, originates from a plateau-like
feature in the magnetization spectral density (see Appendix~\ref{Appendix_A}), which the
Chebyshev expansion resolves accurately. The trivial phases
[Figs.~\ref{fig:Fig_01}(c,d)] exhibit either a vanishing response inside the gap
(electron--hole symmetric case) or a finite plateau when the anti-unitary
particle--hole symmetry is broken, again in quantitative agreement with
${\bf k}$-space calculations.

A notable practical advantage over projector-based real-space implementations is
that our spectral formulation converges efficiently for the magnetization of the
\emph{full} finite sample (including boundary contributions), without requiring
bulk-only averaging over an interior region to obtain stable results. This
provides a direct real-space evaluation of the gauge-invariant total
$M_z(E_F)$, reproducing the full modern-theory magnetization (including the
$\mu\,C$ contribution in Chern insulators) in finite open geometries.

We emphasize that our formulation targets the gauge-invariant total magnetization
of the finite sample. Any real-space ``map'' should be viewed only
as a convention-dependent decomposition of the global trace in Eq. (\ref{eq:2}). In trivial
phases, changing the coordinate origin redistributes the local pattern, modifing the
apparent edge--bulk partition in spatial plots, without
affecting  $M_z$. In the topological insulator case with $\Delta/t_2=0$, \emph{where inside the
bulk gap the orbital magnetization is carried exclusively by edge states},  an origin shift
amounts to an edge--edge redistribution of weight rather than a bulk--edge
repartition (see Fig.~\ref{fig:Sup_03} in the Appendix~\ref{Appendix_B}).

These features highlight that spatial maps of orbital magnetization cannot be
interpreted without accounting for gauge and boundary conventions, and suggest
that sample geometry may play a nontrivial role in itinerant orbital magnetism.
Our real-space spectral method provides a unified and transparent framework for
studying orbital magnetism by treating bulk and boundary responses on equal
footing at the level of the global observable $M_z$. Nevertheless, while not uniquely
defined, such spatial maps remain useful as qualitative diagnostics: in
topological phases they visibly concentrate the in-gap spectral weight near the
boundaries, providing an intuitive real-space signature of bulk--boundary
correspondence (see Appendix~\ref{Appendix_B}).

\begin{figure}[h!]
\centering
\includegraphics[width=1.0\linewidth]{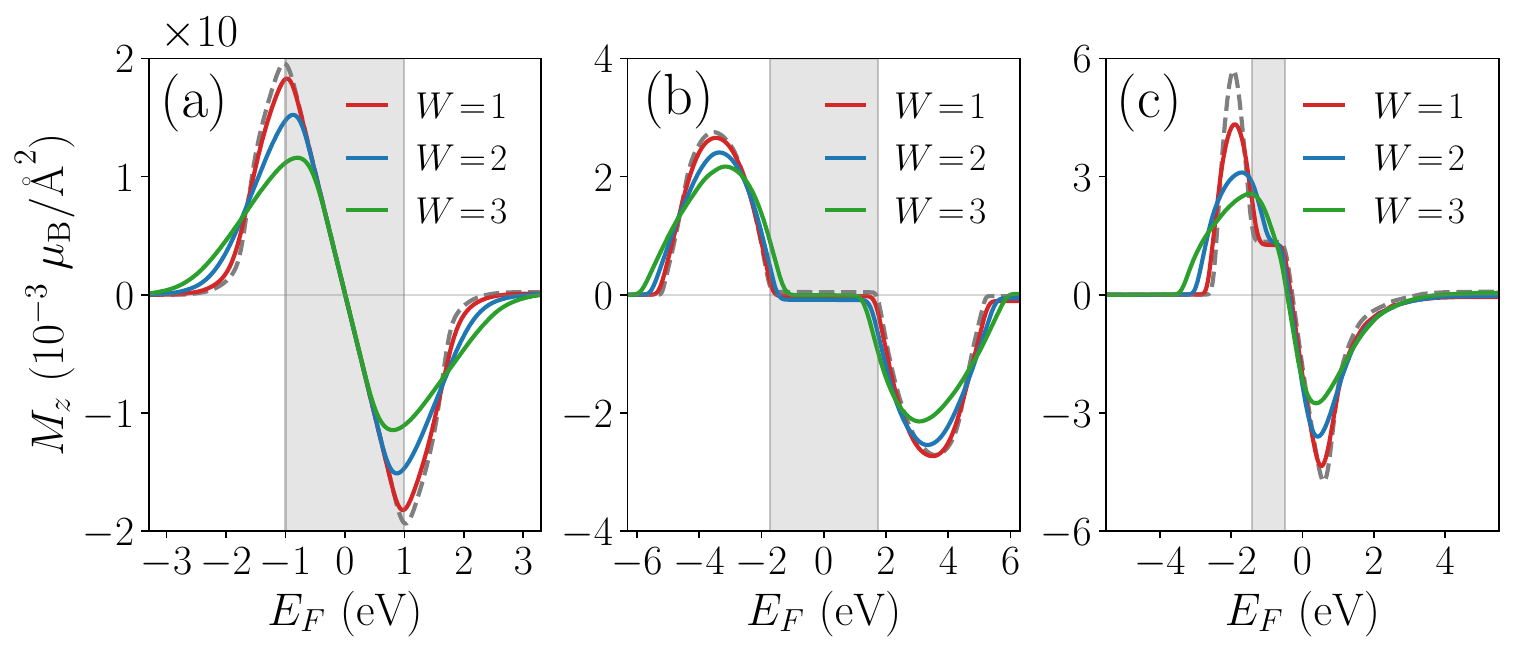}
\caption{
Disorder effects in Haldane model:
(a-c) Orbital magnetization with Anderson disorder as functions of the Fermi energy $E_{F}$ for $W=1$~eV (red), $W=2$~eV (blue) and $W=3$~eV (green). % and $W=3$~eV (orange).
The dashed gray lines correspond to the pristine systems.  
The panels (a), (b), and (c) refer to the red points in $(\phi,\Delta/t_{2})=(0.5\pi,0)$, $(0.5\pi,6\sqrt{3})$, and $(0.1\pi,3)$ of Fig.~\ref{fig:Fig_01}(a).
%The red points in $(\phi,\Delta/t_{2})=(0.5\pi,0)$, $(0.5\pi,6\sqrt{3})$ and $(0.1\pi,3)$ concern the subsequent figures (b), (c) and (d), respectively.
%In our Chebyshev approach, we used a real-space system composed of $9\times 10^{4}$ sites with 200 Chebyshev polynomials and 1000 random vectors, where we use a rectangular geometry.
}
\label{fig:Fig_02}
\end{figure}

For the non-topological phases we consider two representative cases: one with
electron–hole symmetry and one without, corresponding to
Figs.~\ref{fig:Fig_01}(c) and (d).  
In the symmetric insulator ($\phi=\pi/2$, $\Delta/t_2=6\sqrt{3}$), the orbital
magnetization vanishes inside the gap, whereas breaking electron–hole symmetry
($\phi=0.1\pi$, $\Delta/t_2=3$) produces a finite plateau in $M_z(E_F)$.  
In both regimes the magnetization spectral density $m_z(E)$ is zero inside the
gap and only states localized near the physical boundaries contribute to the
global magnetization (see Appendix~\ref{Appendix_A}).  
In the metallic regime the average magnetization receives long-range contributions and a 
gauge-dependent mixture of edge-like and bulk-like character, consistent with the
expected spectral behavior away from a gap.

The utility of our spectral method becomes especially
clear when disorder is introduced.  
Because only sparse matrix–vector multiplications are required to construct the
Chebyshev polynomials, $T_n(\tilde H)$, the total computational cost scales linearly with the
system size and number of Chebyshev iterations.  
This allows unprecedented calculations with up to $10^{10}$ sites—including strongly
disordered or inhomogeneous geometries—without diagonalization or projectors \cite{kite}.
Anderson disorder therefore provides an ideal setting to demonstrate the
strengths of our new approach in large systems of realistic complexity.

We introduce Anderson disorder as $\hat H\rightarrow \hat H+\sum_i \varepsilon_i c_i^\dagger c_i$ with
$\{\varepsilon_i\}$ denoting random on-site energies drawn from a uniform distribution on the interval $[-W/2,W/2]$.  
Figures~\ref{fig:Fig_02}(b–d) show the evolution of $M_z(E_F)$ for increasing
$W$.  
In the topological insulating phase the linear dependence of $M_z(E_F)$ within
the gap remains essentially unchanged as long as the mobility gap persists,
demonstrating the robustness of the boundary spectral flow.  
For the large-gap trivial insulator the magnetization remains zero in the gap
for all $W$, whereas for smaller gaps the plateau shrinks and eventually
disappears when disorder-induced states fill the gap.  {Local magnetization maps further confirm that the space-resolved orbital magnetization, $\langle \hat{\mathscr{M}}_{z}(\textbf{r}) \rangle $, remains extended in the topological case, while trivial phases display isolated, disorder-induced
islands that persist only when the gap is large (see Appendix~\ref{Appendix_B}).  
These results confirm that the orbital magnetization of insulators is a bulk
quantity, even when its real-space distribution is sensitive to gauge choice.

A central conceptual result is the link between the energy derivative of the orbital magnetization and the bulk Chern number, $\mathcal C$ \cite{Bianco2013}. Within our formulation, this link can be expressed in a more direct way through the orbital magnetization spectral density defined in Eq. (\ref{eq:2}):  
\begin{equation}
-\frac{2\hbar c}{e}\,m_{z}(E_{F})=\frac{\mathcal{C}}{2\pi},    
\end{equation}
where $\mathcal C$ denotes the Chern number of occupied states. Figure~\ref{fig:Fig_03}(a) shows $-m_{z}(E_{F})=-dM_z/dE_F$ for several parameter values near
the $\mathcal{C}=0\to 1$ transition of the Haldane model.  
We use this property to test the method in the vicinity of the topological
transition of the Haldane model.  We select several points near the
$\mathcal{C}=0\!\to\!1$ phase boundary at fixed $\Delta/t_2=3\sqrt{3}\sin(\pi/4)$ and varying $\phi$, and
evaluate $m_z$ at the charge-neutrality point (CNP),  to ensure that $E_F$ lies
within the gap [Fig.~\ref{fig:Fig_03}(a)]. In the thermodynamic limit the transition occurs at
$\phi=0.25\pi$; in finite systems the change is necessarily smooth, and our
results follow this expected behavior. The plateaus reproduce the expected integer Chern numbers, and the transition sharpens with increasing system size [Fig.~\ref{fig:Fig_03}(b)], as expected in
finite-size topological systems.   For $\phi < 0.25\pi$ we obtain
$\mathcal{C}\simeq 0$, while for $\phi> 0.25\pi$ we recover $\mathcal{C}\simeq 1$.  As the
system size increases, the crossover sharpens accordingly. Importantly, the Chern number is extracted here without projectors, curvature markers, Wannier functions, or assumptions about translational symmetry, in
contrast with existing real-space approaches~\cite{Kitaev2006,He2019,Chen2023}.

In the presence of disorder [Fig.~\ref{fig:Fig_03}(c)], the plateaus broaden, however, the method can still distinguish the two different topological phases for systems with well defined gaps. Disorder also helps to converge the Chebyshev expansion to the $C$ of the thermodynamic limit at the critical point of the topological phase transition for finite systems. Panel Fig.~\ref{fig:Fig_03}(d) shows the convergence of the Chern number as a function of the number of Chebyshev moments $M$ in the expansion.

\begin{figure}[h!]
\centering
\includegraphics[width=1.0\linewidth]{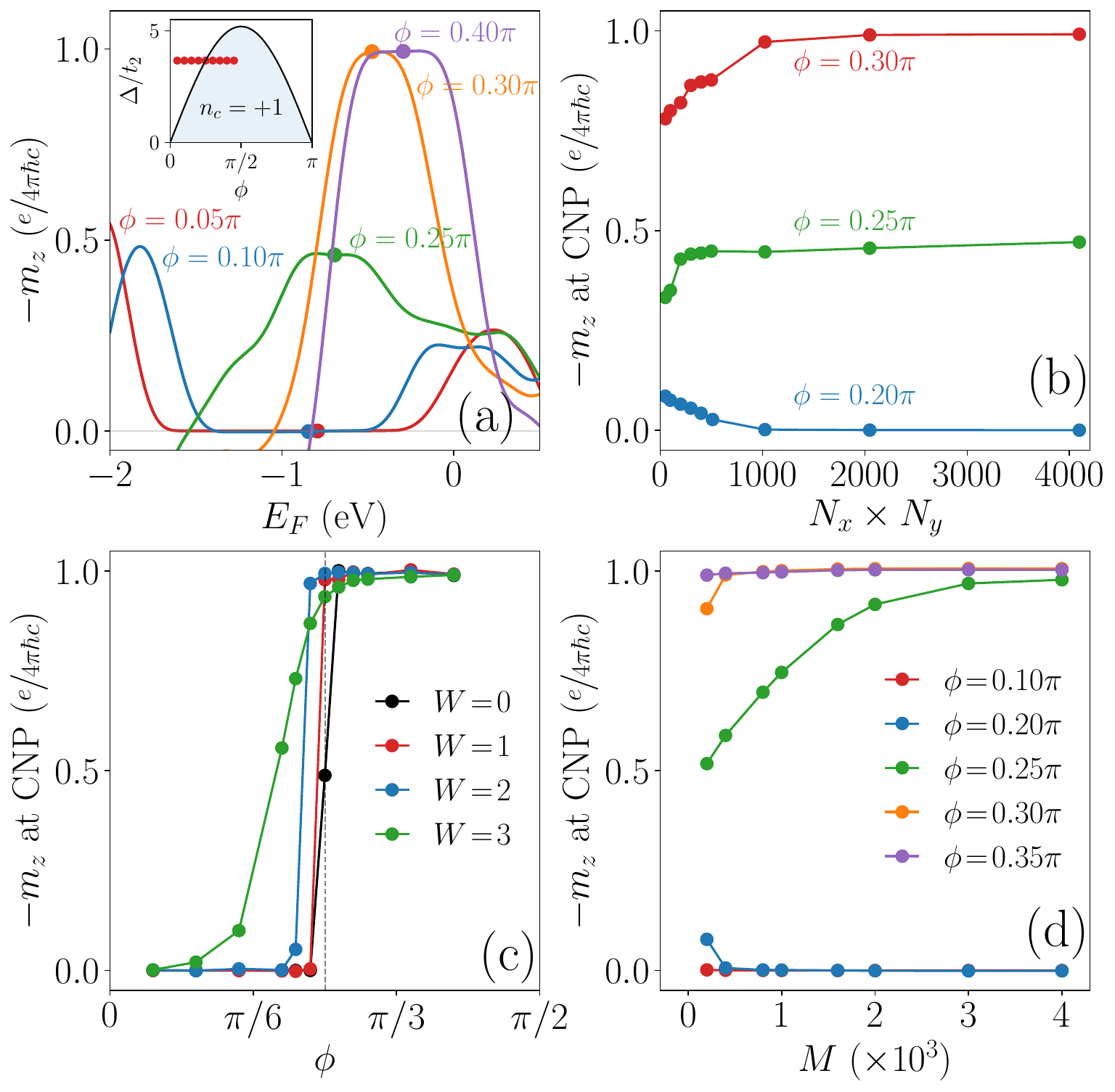}
\caption{
Chern number from the orbital magnetization:
Panel (a) plots \(-m_z\), calculated within the band gap for pristine systems at \(\Delta/t_2=3\sqrt{3}\sin(\pi/4)\) and \(\phi= 0.05\pi\), \(0.10\pi\), \(0.25\pi\), \(0.30\pi\) and \(0.40\pi\). The colored dots on each curve denote the respective CNP values.
Other simulation parameters: $D\approx4\times10^6$, $M=200$ and $R=2000$.
The inset represents the Chern number of the valence band of the Haldane model. 
The red points in this phase diagram are relevant for the subsequent discussion and figures.
Panel (b) plots \(-m_z\) at the CNP as a function of the sample sizes for pristine systems with \(\phi= 0.20\pi\), \(0.25\pi\) and \(0.30\pi\).
The panel (c) plots \(-m_z\) at the CNP as functions of \(\phi\) with $D\approx4\times10^6$, $M=4000$ and $R=2000$.
The panel (d) plots \(-m_z\) at the CNP as functions of \(M\) for \(W=1\) with $D\approx4\times10^6$ and $R=2000$.
%The panels (c) and (d) plot \(-m_z\) at the CNP as functions of \(\phi\) and \(M\).
}
\label{fig:Fig_03}
\end{figure}

\begin{figure}[h!]
\centering
\includegraphics[width=1.0\linewidth]{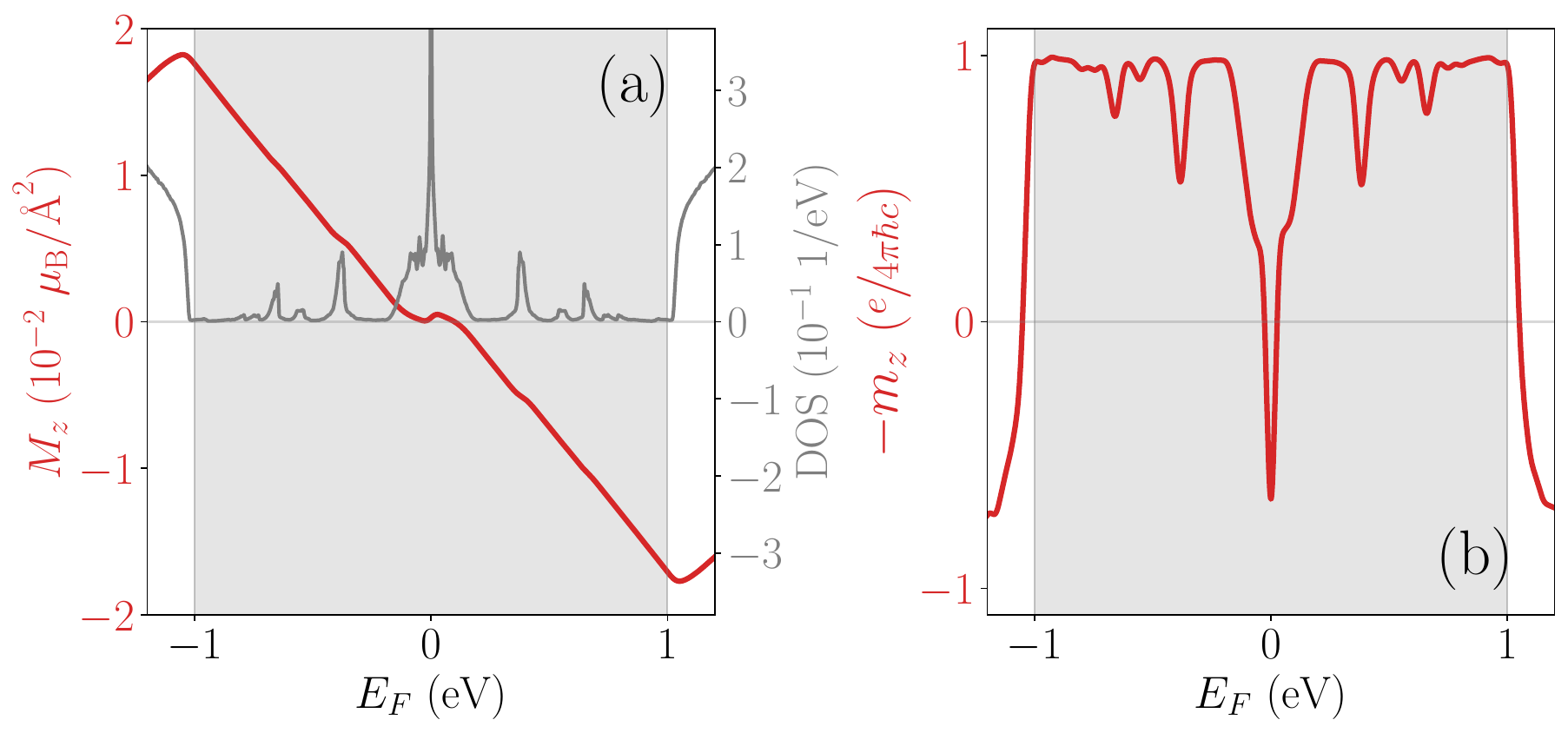}
\caption{
Point defects in the Haldane model:
Panel (a) plots the orbital magnetization (red line) in real-space formulation and density of states (gray line) as functions of the energy, within the band gap for the topological insulating phase with \(4\%\) vacancy concentration, \(\phi=0.5\pi\) and \(\Delta/t_2=0\).
Panel (b) plots \(-m_z\), calculated within the band gap for the same topological insulating phase. Other simulation parameters: $D\approx10^6$, $M=800$ and $R=2400$.
}
\label{fig:Fig_04}
\end{figure}

We turn now to vacancy defects, which generate qualitatively different physics
from Anderson disorder \cite{Pires2022}.  Vacancies produce
mid-gap resonances whose structure depends sensitively on particle–hole symmetry \cite{Ferreira2015}
and on whether the system is topological or trivial.  In topological insulators on
the honeycomb lattice, single vacancies are known to bind zero or near-zero modes
with wave functions strongly peaked around the defect and hybridized with the
chiral edge manifold \cite{He2013}.  These resonances do not produce
in-gap extended states, but instead appear as sharp peaks near the band edges or
at $E=0$ in systems with approximate chiral symmetry, consistent with our density
of states in Fig.~\ref{fig:Fig_04}(a).  Their spectral weight originates from the
same subspace responsible for the boundary contribution to the orbital
magnetization, and therefore affects the linear dependence of $M_z(E_F)$ in a
very characteristic way.

To explore this regime, we consider a vacancy concentration of $4\%$ in a
rectangular flake with $10^6$ lattice sites.  In the pristine topological phase
($\phi=\pi/2$, $\Delta/t_2=0$), the orbital magnetization displays a perfectly
linear dependence inside the bulk gap.  With vacancies, this linearity survives
almost entirely: it is disrupted only in narrow energy windows centered at the
vacancy-induced resonances.  Each in-gap (or near-edge) peak in the density of
states produces a corresponding kink or inflection in $M_z(E_F)$, reflecting
the sensitivity of the boundary magnetization to the sharp redistribution of
spectral weight.  The resonance near $E=0$ generates the strongest deviation,
forming an odd-symmetric feature reminiscent of the zero-mode physics of
graphene vacancies.

Surprisingly, the global trend of $M_z$—including the overall slope associated with
the Chern number—remains intact even in the presence of relatively dense vacancy disorder, while the global topology encoded in the average Chern number is strongly affected near the resonances [see Fig. \ref{fig:Fig_04}(b)]. 
This indicates that (i) the orbital
magnetization is robust to vacancy scattering, and (ii) the vacancy resonances,
although strongly localized, couple only weakly to the long-range orbital
circulation that determines the macroscopic magnetization.  As a consequence, the
Chebyshev-based real-space approach can resolve both types of contributions
simultaneously: the sharp, defect-induced resonances and the robust topological
linear background.

In summary, we introduced a fully real-space formulation of orbital
magnetization based on the Chebyshev expansion of the spectral operator
$\hat{\mathscr{M}}_{z}\,\delta(E-\hat H)$.  
This method avoids projectors and eigenstates entirely, works naturally with
open boundaries, and scales linearly with system size, enabling simulations investigations of orbital magnetization in systems far more realistic and larger than those accessible to existing approaches. Because the formulation remains valid in the presence of disorder and
spatial inhomogeneities, it captures the evolution of orbital magnetization under uncorrelated on-site disorder and vacancy-induced resonances, revealing the conditions under which the topological signatures of the orbital magnetization can be considered to be robust. These results establish the Chebyshev spectral approach as a compact, computationally cheap, and broadly applicable framework for orbital magnetization in large, disordered, and topological systems.

\begin{acknowledgments}
TGR, JMVPL and AF acknowledge the  FCT - Fundação para a Ciência e Tecnologia for financial support (Grant No. 2023.11755.PEX and DOI identifier https://doi.org/10.54499/2023.11755.PEX)
TGR acknowledges  FCT  project reference numbers UIDB/04650/2020,  and 2022.07471.CEECIND/CP1718/CT0001
(with DOI identifier: 10.54499/2022.07471.CEECIND/CP1718/
CT0001) and CNPq (Grant No 305013/2024-6). TGR and KJUV acknowledge support from the EIC Pathfinder OPEN grant 101129641 “OBELIX”. TPC acknowledges CNPq (Grant No. 305647/2024-5). HPV, JMVPL and AF acknowledge the University of York High-Performance Computing service, Viking.
Further support from FCT through the PhD Grant. No. 2024.00560.BD (HPV) is acknowledged.

\end{acknowledgments}

\appendix

\section{Orbital magnetization spectral density} 
\label{Appendix_A}
\setcounter{figure}{0}
\renewcommand{\thefigure}{A\arabic{figure}}

The orbital magnetization spectral density is defined as \(m_z(E)=\,\mathrm{Tr}\big[\hat{\mathscr{M}}_{z}\,\delta(E-\hat H)\big]\). 
To evaluate this quantity efficiently, we expand the Dirac-delta operator in terms of Chebyshev polynomials of the Hamiltonian matrix. Technically, this requires first rescaling the Hamiltonian and energy variable according to \(\hat H\!\to\!\tilde H\) and \(E\!\to\!\tilde E\), ensuring that the dimensionless energy falls within the canonical interval, $\mathcal I = [-1,1]$, of orthogonal polynomial expansions. This approach involves computing moments of the resulting Chebyshev expansion.  The total orbital magnetization \(M_z(E_F)\) is finally obtained by integrating  \(m_z(E)\) up to the Fermi energy \(E_F\), as outlined in the main text. In Fig.~\ref{fig:Sup_01}, we consider the three representative cases addressed in the main text. Panels (a), (b) and (c) of Fig. \ref{fig:Sup_01} use, respectively, the set of parameters of panels (b), (c) and (d) of Fig.~1 (main text).

\begin{figure*}%[h!]
\centering
\includegraphics[width=0.80\linewidth]{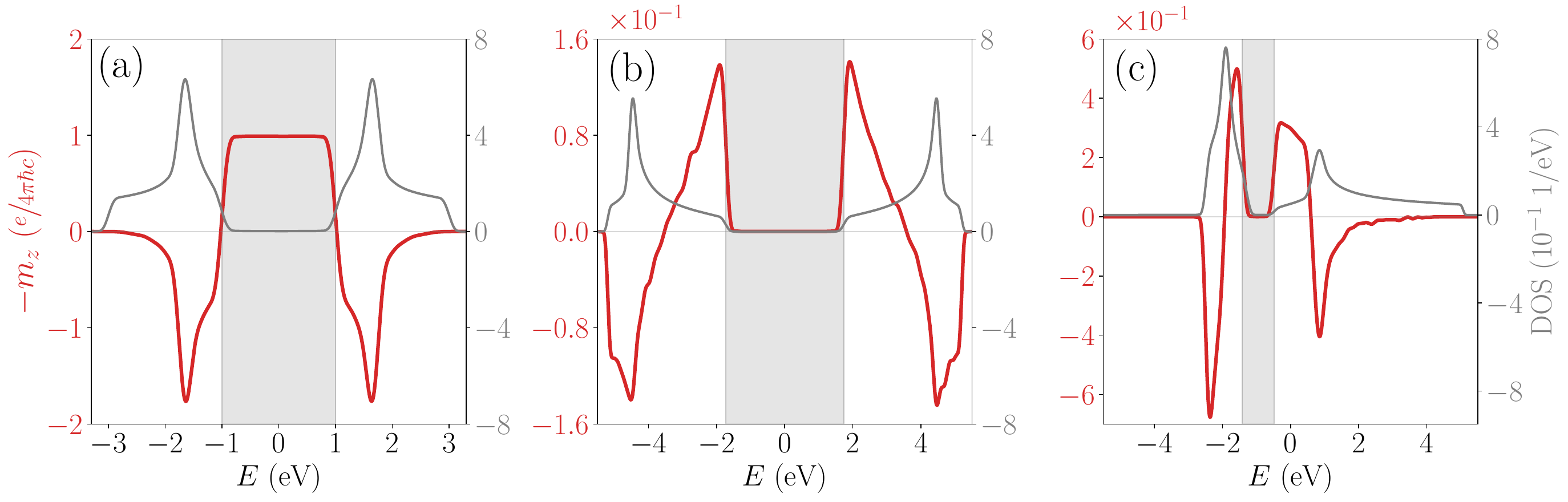}
\caption{
(a-c) Orbital magnetization spectral density (red lines) in real-space formulation and density of states (gray lines) as functions of the energy. Panels (a), (b), and (c) show results for the systems with $(\phi,\Delta/t_{2})=(0.5\pi,0)$, $(0.5\pi,6\sqrt{3})$, and $(0.1\pi,3)$, respectively. Calculations are performed on a rectangular domain with $10^{6}$ sites. Chebyshev parameters: $M=200$ and $R=800$.}
\label{fig:Sup_01}
\end{figure*}

\begin{figure*}%[h!]
\centering
\includegraphics[width=0.76\linewidth]{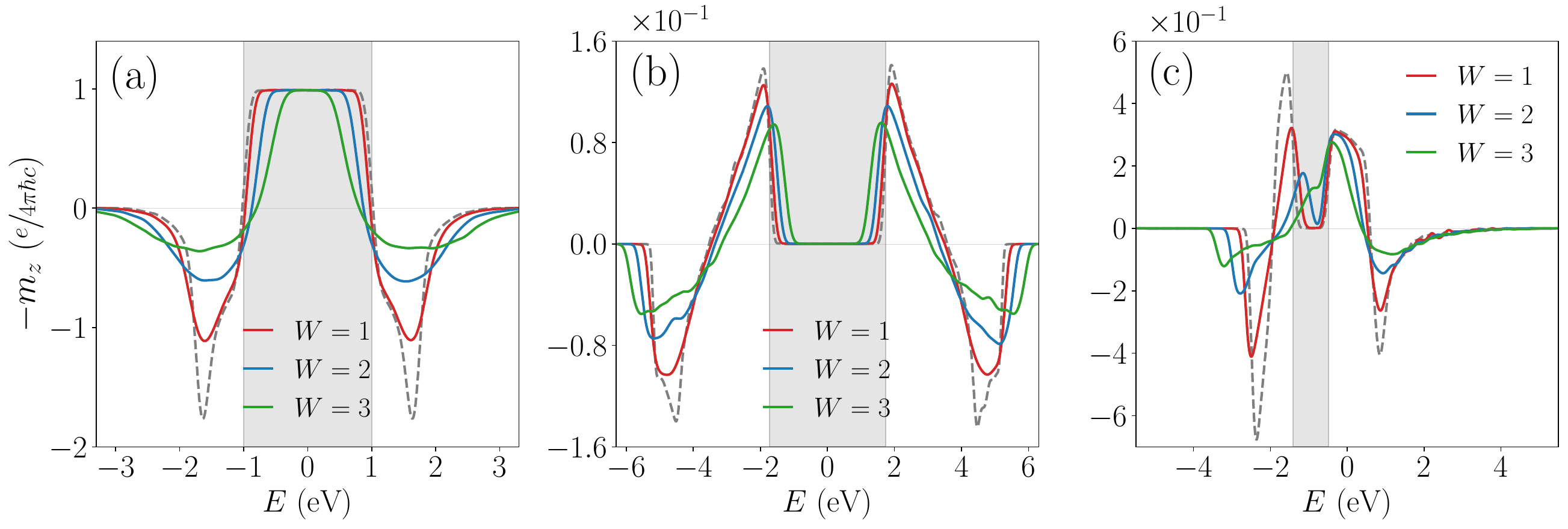}
\caption{
Disorder effects on the Haldane model:
The panels (a), (b) and (c) concern the red points in $(\phi,\Delta/t_{2})=(0.5\pi,0)$, $(0.5\pi,6\sqrt{3})$ and $(0.1\pi,3)$ of the Fig.~1(a), respectively.
(a-c) Energy dependence of orbital magnetization spectral density for selected Anderson disorder strengths: 
$W=1$~eV (red), $W=2$~eV (blue) and $W=3$~eV (green). % and $W=3$~eV (orange).
The dashed gray lines correspond to the pristine systems.  
Simulation parameters as in Fig. \ref{fig:Sup_01}.
}
\label{fig:Sup_02}
\end{figure*}

For the topological insulating phase with \(\phi=0.5\pi\) and \(\Delta/t_2=0\), corresponding to Fig.~\ref{fig:Sup_01}(a), the orbital magnetization spectral density exhibits a well-defined plateau within the gap. This plateau has a sign opposite to that of \(m_z(E)\) in the metallic regime.
In contrast, for the non-topological insulating phases shown in Figs.~\ref{fig:Sup_01}(b) and (c), the orbital magnetization spectral density vanishes within the gap.
In particular, phases with \(\phi\neq 0.5\pi\) have broken particle-hole symmetry and, consequently, integrated spectral function over the valence band yields a non-zero orbital magnetization; see Fig.~\ref{fig:Sup_01}(c) together with Fig.~1(d) of the main text.

Figures~\ref{fig:Sup_02} show that weak Anderson disorder broadens the features of energy-resolved orbital magnetization spectral density as expected. Thus, the smaller the gap, the less robust the features are to disorder. When the disorder strength exceeds the (pristine) energy gap scale, the impact of disorder is quite significant [see e.g. Fig.~\ref{fig:Sup_02} (c)].

The real-space approach based on accurate Chebyshev polynomial expansions enables the calculation of the global Chern number via the derivative of the orbital magnetization with respect to the Fermi energy, \(\dv{M_z}{E_F}\). 
By construction, this quantity is the energy-resolved orbital magnetization spectral density, $m_z(E_F)$. 
Hence, $m_z(E_F)$ itself encodes information about the global Chern number of the system.
Specifically, its value in the spectral gap is directly proportional to the Chern number of occupied states, \(m_z(E_F=0) = -\left( \frac{e}{2\hbar c} \right) \frac{\mathcal{C}}{2\pi} \).
Therefore, for a Chern number \(\mathcal{C}=1\), the rescaled magnetization spectral density has a value of approximately \(-0.02\;\mu_{\rm B}/\)eV{\AA}\(^2\) across the band gap.

\section{Spectral function of the local atomic environment} 
\label{Appendix_B}
\setcounter{figure}{0}
\renewcommand{\thefigure}{B\arabic{figure}}

We now examine the spatial structure of the orbital magnetization in real space.
We begin with clean samples and later introduce disorder.  The Hamiltonian
parameters correspond to the three representative cases marked by red points in
Fig.~1(a) of the main text.

Figure~\ref{fig:Sup_03}(a) shows the finite flake used in this analysis: a
rectangular geometry with 100 sites along the zigzag direction and 200 sites
along the armchair direction.  To illustrate the origin dependence of local
quantities involving the position operator, we compute the spatially resolved
expectation value
$\langle \hat{\mathscr M}_z(\mathbf r)\rangle$ for a clean topological sample
with $\Delta/t_2=0$ and $\phi=\pi/2$.
\begin{figure*}%[h!]
\centering
\includegraphics[width=0.83\linewidth]{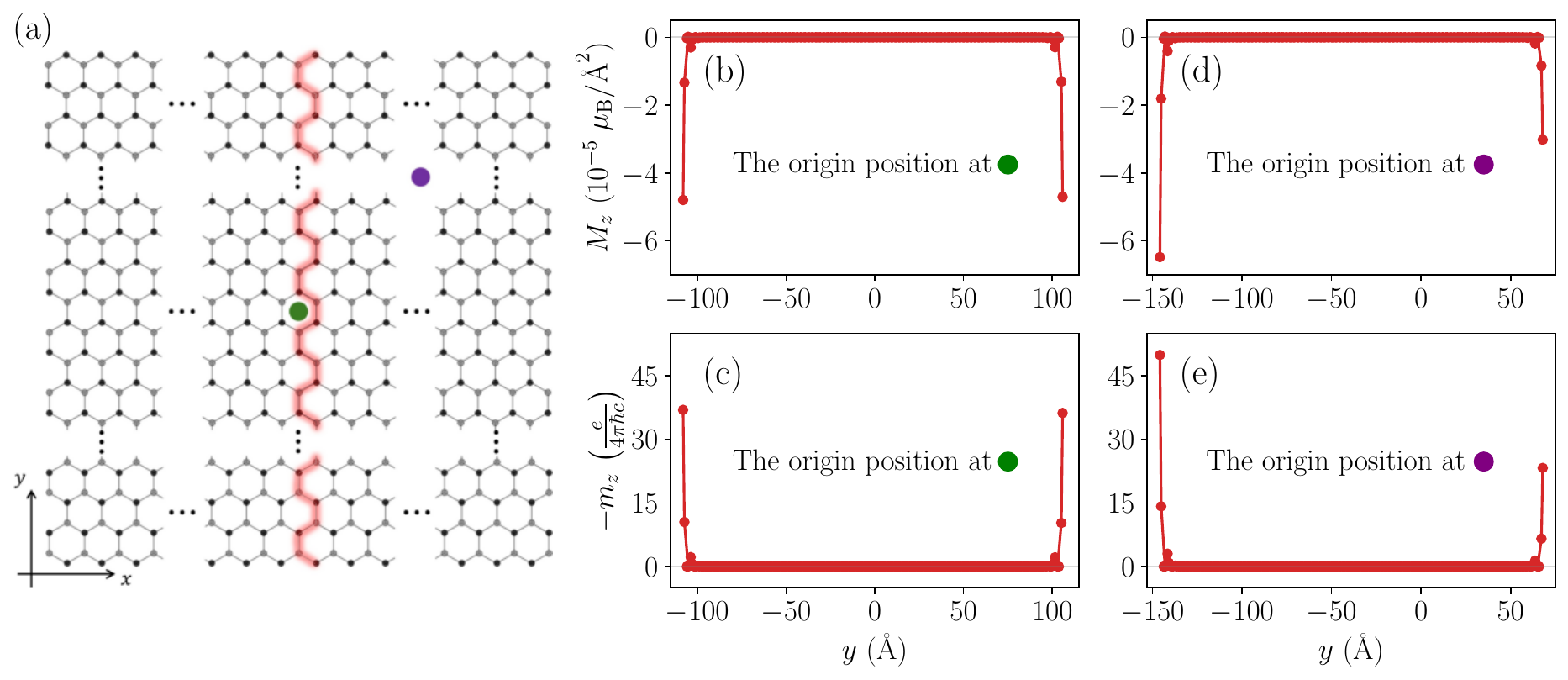}
\caption{(a) A flake with a rectangular geometry, composed of \(1\times10^4\) sites, representing the honeycomb lattice of the Haldane model. 
Black and grey circles indicate nonequivalent sites. 
(b)-(e) One-dimensional plots of the local orbital magnetization and local Chern number at \(E_F=0\)~eV for the 200 sites along the vertical red line shown in (a). Here we set \(\Delta/t_2 =0\) and \(\phi=0.5\pi\), which corresponds to the topological insulating phase.
The coordinate origin location chosen in each panel (b)-(e) is marked by the green [purple] point in panel (a).}
\label{fig:Sup_03}
\end{figure*}

Panels (b) and (d) of Fig.~\ref{fig:Sup_03} show the profile of
$\langle \hat{\mathscr M}_z(\mathbf r)\rangle$ along the vertical line
highlighted in panel (a).  When the coordinate origin is chosen at the geometric
center (green point), the two edges display identical contributions, as required
by symmetry.  When the origin is shifted away from the center (purple point), the
two edges become inequivalent.  This dependence is expected: because
$\hat{\mathscr M}_z$ contains the position operators explicitly,
$\langle \hat{\mathscr M}_z(\mathbf r)\rangle$ is origin dependent, while the
spatial average defining the total orbital magnetization is completely
origin independent.

Panels (c) and (e) show the corresponding \textit{local orbital magnetization
spectral density}.  This quantity also varies with the coordinate origin, again
reflecting the gauge dependence inherent to real-space expressions containing
the position operator.  For the topological insulating case, the spectral density
vanishes in the bulk and acquires significant weight at the edges.  Its spatial
average is quantized and equal to the global Chern number, whereas for the
non-topological insulating phase the spectral density vanishes everywhere.

It is useful to contrast this behavior with the construction of the local
topological marker introduced by Bianco and Resta~\cite{Resta2011}.  Their
marker is engineered to be a bulk quantity even in finite samples: the insertion
of $\mathscr{P}$ and $\mathscr{Q}=1-\mathscr{P}$ suppresses the matrix elements of the position operator that
connect bulk and edge subspaces.  As a consequence, the local marker is well
defined and convergent only in the interior of a large sample; near the physical
boundaries it becomes large and eventually diverges, reflecting the fact that it
is not intended to describe edge physics.  In our case, however, the edge
contributions are \emph{not} pathologies to be projected out.  For orbital
magnetization in a finite Chern insulator, the boundary states carry the spectral
flow responsible for the linear dependence of $M_z(E_F)$ observed in the main
text.  The quantity plotted in Fig.~\ref{fig:Sup_03} therefore represents the
actual spatial distribution of the orbital magnetization spectral density, not a
local topological invariant.

\begin{figure*}%[h!]
\centering
\includegraphics[width=0.70\linewidth]{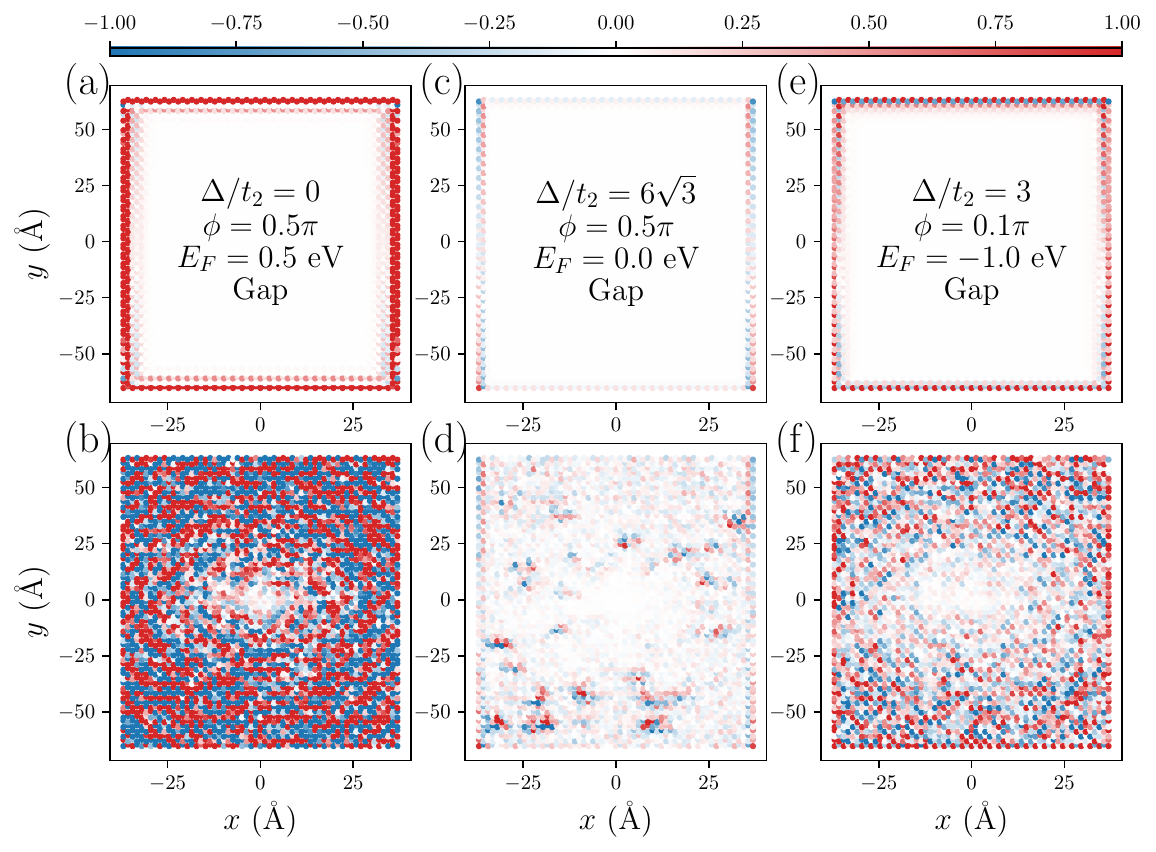}
\caption{
Local orbital magnetization, displaying only values in the range $[-1, 1]\;\mu_{\rm B}$, in real space for a system composed of $3.6\times 10^{3}$ sites with a rectangular geometry.
First row: pristine systems. Second row: systems with Anderson disorder ($W=2$~eV).
The panels (a) and (b) correspond to the topological insulating phase $(\phi=0.5\pi, \Delta/t_{2}=0)$ with $E_{F}= -0.5$ eV and $0$, respectively; the panels (c) and (d) correspond to the non-topological insulating phase $(\phi=0.5\pi, \Delta/t_{2}=6\sqrt{3})$ with $E_{F}= 0$; and the panels (e) and (f) correspond to the non-topological insulating phase $(\phi=0.1\pi, \Delta/t_{2}=3)$ with $E_{F}= -1$ eV. 
The red (blue) points correspond to positive (negative) orbital magnetization.
We used 200 Chebyshev polynomials.}
\label{fig:Sup_04}
\end{figure*}

We now turn to the effect of Anderson disorder, complementing the results of the
main text.  Figure~\ref{fig:Sup_04}(b) shows that in the topological insulating
phase the orbital magnetization remains delocalized throughout the sample and is
remarkably robust against disorder.  In contrast, for the trivial insulating
phase with a large gap, disorder induces isolated mid-gap states that manifest
as small, spatially localized patches of orbital magnetization [panel (d)].  When
the disorder strength becomes comparable to the gap, these localized regions
expand and the orbital magnetization spreads across the sample [panel (f)],
consistent with a transition toward an Anderson insulating or critical regime.

These results illustrate how disorder modifies the spatial structure of the
orbital magnetization in distinct ways depending on the underlying phase, while
preserving the global behavior of $M_z$ discussed in the main text.

\section{Spectral function of the local atomic environment} 
\label{Appendix_C}
\setcounter{figure}{0}
\renewcommand{\thefigure}{C\arabic{figure}}

This section further showcases our spectral real-space approach by presenting additional representative cases and comparing the results against those derived from the reciprocal space formula of the modern theory of orbital magnetization. 
These comparisons serve to underscore the computational practicality and effectiveness of our real-space approach to calculate this fundamental property in crystalline materials.

Figure~\ref{fig:Sup_05}(b) compares our real-space results, obtained using the Chebyshev expansion for \(\Delta/t_2 =3\) across the range \(\phi=0\) to \(\pi\), with two established methods: the modern theory (bulk formula) in reciprocal space and the extrapolated results from finite-sample calculations (heuristic formula) of Fig. 7 in Ref.~\cite{Resta2006}. 
The results show excellent agreement with our real-space approach based on Chebyshev polynomials.

\begin{figure*}%[h!]
\centering
\includegraphics[width=0.55\linewidth]{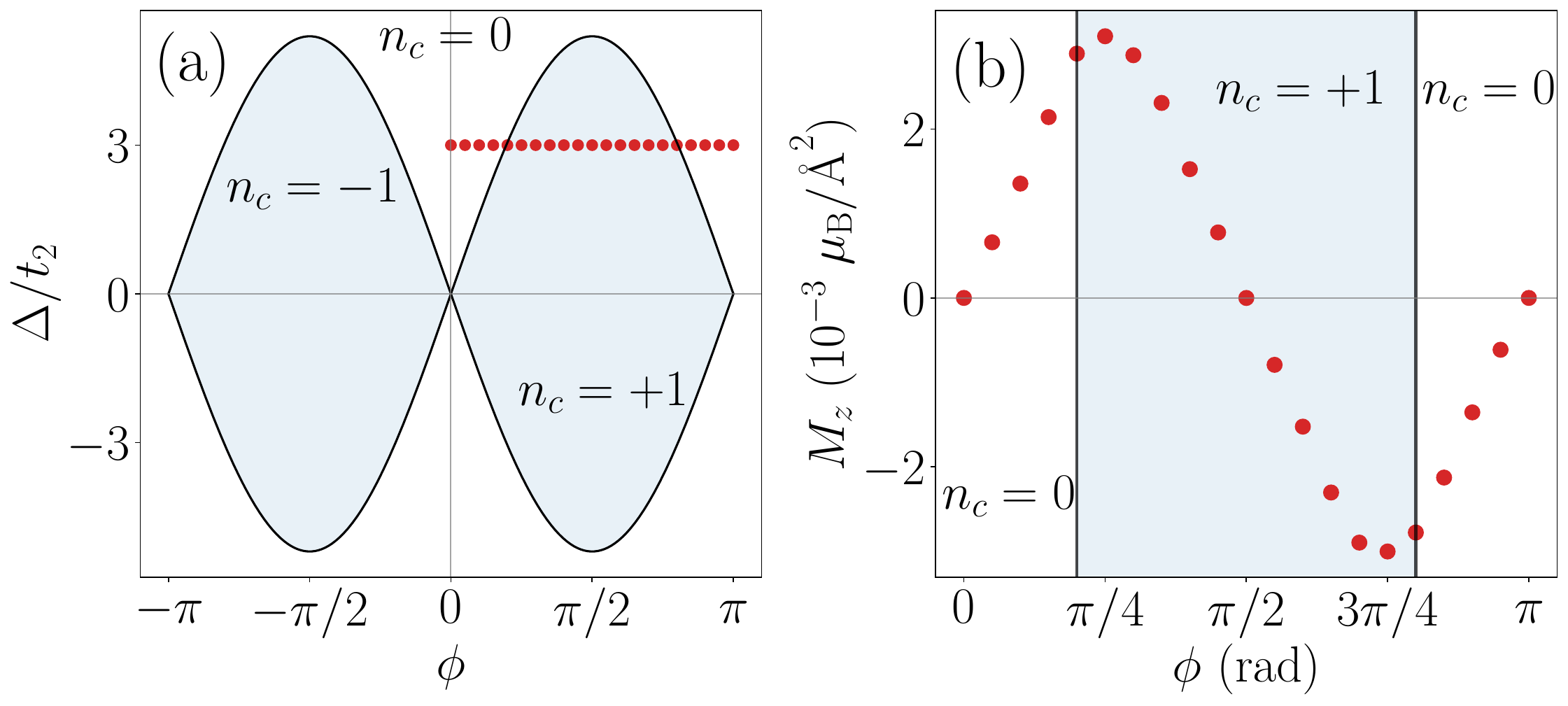}
\caption{
Panel (a) illustrated the Chern number of the valence band of the Haldane model as a function of the paramenters \(\phi\) and \(\Delta/t_2\) (\(t_1=2\)~eV and \(t_2=1/3\)~eV).
The red points concern in the subsequent figure.
(b) Orbital magnetization of the Haldane model with  \(\Delta/t_2=3\) as a function of the paremeter \(\phi\), where the lowest band is treated as occupied.
The system has nonzero Chern number in the region between two line vertical lines.
}
\label{fig:Sup_05}
\end{figure*}

Figures~\ref{fig:Sup_06}(b)-(l) show plots of the Fermi energy dependence of the orbital magnetization for \(\Delta/t_2 =3\) with \(\phi=0\), \(0.1\pi\), \(0.2\pi\), \(0.3\pi\), \(0.4\pi\), \(0.5\pi\), \(0.6\pi\), \(0.7\pi\), \(0.8\pi\), \(0.9\pi\) e \(\pi\), respectively, using our real-space formalism (red lines) and the reciprocal-space formalism (blue points).
We observe that there are pairs of identical curves using \(\phi=0.5\pi\) as a reference point. 
For example, the pair of curves for \(\phi=0.3\pi\) and \(\phi=0.7\pi\) are identical but exhibit inverted energy bands and a sign reversal in orbital magnetization.

\begin{figure*}%[h]
\centering
\includegraphics[width=0.85\linewidth]{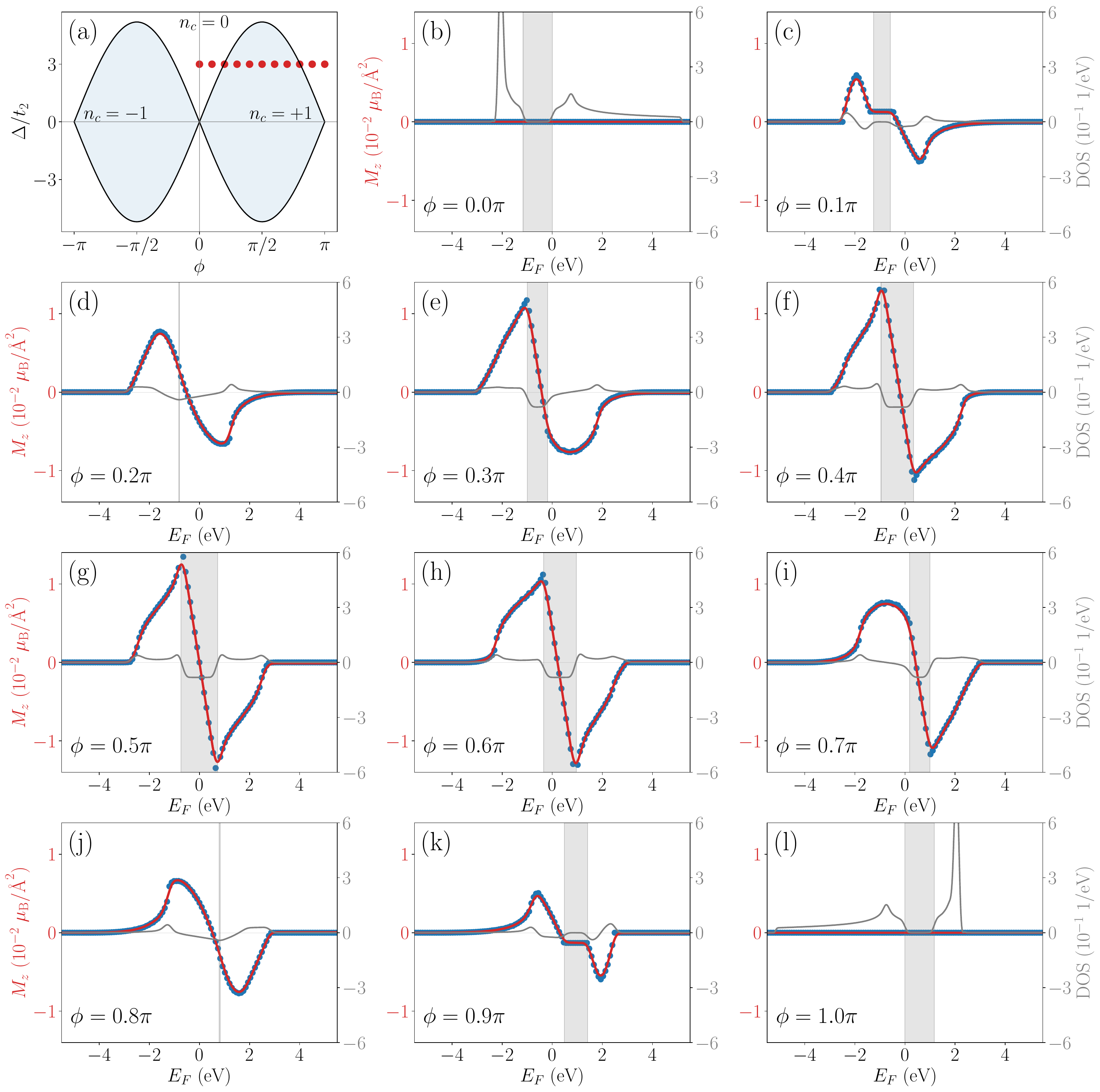}
\caption{
Panel (a) illustrates the Chern number of the bottom band of the Haldane model as a function of the paramenters \(\phi\) and \(\Delta/t_2\) (\(t_1=2\)~eV and \(t_2=1/3\)~eV).
The red points concern the subsequent figures.
(b)-(l) Orbital magnetization (red lines) in real-space formulation and density of states (gray lines) as functions of the energy, \(E_F\) and \(E\), respectively.
The blue points correspond to the orbital magnetization in the reciprocal-space formulation.
In our spectral simulations, we used a real-space system composed of \(10^6\) sites (rectangular geometry) with 200 Chebyshev polynomials and 800 random vectors.}
\label{fig:Sup_06}
\end{figure*}

\bibliography{ref}

\begin{thebibliography}{49}%
\makeatletter
\providecommand \@ifxundefined [1]{%
 \@ifx{#1\undefined}
}%
\providecommand \@ifnum [1]{%
 \ifnum #1\expandafter \@firstoftwo
 \else \expandafter \@secondoftwo
 \fi
}%
\providecommand \@ifx [1]{%
 \ifx #1\expandafter \@firstoftwo
 \else \expandafter \@secondoftwo
 \fi
}%
\providecommand \natexlab [1]{#1}%
\providecommand \enquote  [1]{``#1''}%
\providecommand \bibnamefont  [1]{#1}%
\providecommand \bibfnamefont [1]{#1}%
\providecommand \citenamefont [1]{#1}%
\providecommand \href@noop [0]{\@secondoftwo}%
\providecommand \href [0]{\begingroup \@sanitize@url \@href}%
\providecommand \@href[1]{\@@startlink{#1}\@@href}%
\providecommand \@@href[1]{\endgroup#1\@@endlink}%
\providecommand \@sanitize@url [0]{\catcode `\\12\catcode `\$12\catcode
  `\&12\catcode `\#12\catcode `\^12\catcode `\_12\catcode `\%12\relax}%
\providecommand \@@startlink[1]{}%
\providecommand \@@endlink[0]{}%
\providecommand \url  [0]{\begingroup\@sanitize@url \@url }%
\providecommand \@url [1]{\endgroup\@href {#1}{\urlprefix }}%
\providecommand \urlprefix  [0]{URL }%
\providecommand \Eprint [0]{\href }%
\providecommand \doibase [0]{https://doi.org/}%
\providecommand \selectlanguage [0]{\@gobble}%
\providecommand \bibinfo  [0]{\@secondoftwo}%
\providecommand \bibfield  [0]{\@secondoftwo}%
\providecommand \translation [1]{[#1]}%
\providecommand \BibitemOpen [0]{}%
\providecommand \bibitemStop [0]{}%
\providecommand \bibitemNoStop [0]{.\EOS\space}%
\providecommand \EOS [0]{\spacefactor3000\relax}%
\providecommand \BibitemShut  [1]{\csname bibitem#1\endcsname}%
\let\auto@bib@innerbib\@empty
%</preamble>
\bibitem [{\citenamefont {Sharpe}\ \emph {et~al.}(2019)\citenamefont {Sharpe},
  \citenamefont {Fox}, \citenamefont {Barnard}, \citenamefont {Finney},
  \citenamefont {Watanabe}, \citenamefont {Taniguchi}, \citenamefont
  {Kastner},\ and\ \citenamefont {Goldhaber-Gordon}}]{Aaron2019}%
  \BibitemOpen
  \bibfield  {author} {\bibinfo {author} {\bibfnamefont {A.~L.}\ \bibnamefont
  {Sharpe}}, \bibinfo {author} {\bibfnamefont {E.~J.}\ \bibnamefont {Fox}},
  \bibinfo {author} {\bibfnamefont {A.~W.}\ \bibnamefont {Barnard}}, \bibinfo
  {author} {\bibfnamefont {J.}~\bibnamefont {Finney}}, \bibinfo {author}
  {\bibfnamefont {K.}~\bibnamefont {Watanabe}}, \bibinfo {author}
  {\bibfnamefont {T.}~\bibnamefont {Taniguchi}}, \bibinfo {author}
  {\bibfnamefont {M.~A.}\ \bibnamefont {Kastner}},\ and\ \bibinfo {author}
  {\bibfnamefont {D.}~\bibnamefont {Goldhaber-Gordon}},\ }\bibfield  {title}
  {\bibinfo {title} {Emergent ferromagnetism near three-quarters filling in
  twisted bilayer graphene},\ }\href {https://doi.org/10.1126/science.aaw3780}
  {\bibfield  {journal} {\bibinfo  {journal} {Science}\ }\textbf {\bibinfo
  {volume} {365}},\ \bibinfo {pages} {605} (\bibinfo {year}
  {2019})}\BibitemShut {NoStop}%
\bibitem [{\citenamefont {Liu}\ \emph {et~al.}(2019)\citenamefont {Liu},
  \citenamefont {Ma}, \citenamefont {Gao},\ and\ \citenamefont
  {Dai}}]{Liu2019}%
  \BibitemOpen
  \bibfield  {author} {\bibinfo {author} {\bibfnamefont {J.}~\bibnamefont
  {Liu}}, \bibinfo {author} {\bibfnamefont {Z.}~\bibnamefont {Ma}}, \bibinfo
  {author} {\bibfnamefont {J.}~\bibnamefont {Gao}},\ and\ \bibinfo {author}
  {\bibfnamefont {X.}~\bibnamefont {Dai}},\ }\bibfield  {title} {\bibinfo
  {title} {Quantum valley hall effect, orbital magnetism, and anomalous hall
  effect in twisted multilayer graphene systems},\ }\href
  {https://doi.org/10.1103/PhysRevX.9.031021} {\bibfield  {journal} {\bibinfo
  {journal} {Phys. Rev. X}\ }\textbf {\bibinfo {volume} {9}},\ \bibinfo {pages}
  {031021} (\bibinfo {year} {2019})}\BibitemShut {NoStop}%
\bibitem [{\citenamefont {Tschirhart}\ \emph {et~al.}(2021)\citenamefont
  {Tschirhart}, \citenamefont {Serlin}, \citenamefont {Polshyn}, \citenamefont
  {Shragai}, \citenamefont {Xia}, \citenamefont {Zhu}, \citenamefont {Zhang},
  \citenamefont {Watanabe}, \citenamefont {Taniguchi}, \citenamefont {Huber},\
  and\ \citenamefont {Young}}]{Tschirhart2021}%
  \BibitemOpen
  \bibfield  {author} {\bibinfo {author} {\bibfnamefont {C.~L.}\ \bibnamefont
  {Tschirhart}}, \bibinfo {author} {\bibfnamefont {M.}~\bibnamefont {Serlin}},
  \bibinfo {author} {\bibfnamefont {H.}~\bibnamefont {Polshyn}}, \bibinfo
  {author} {\bibfnamefont {A.}~\bibnamefont {Shragai}}, \bibinfo {author}
  {\bibfnamefont {Z.}~\bibnamefont {Xia}}, \bibinfo {author} {\bibfnamefont
  {J.}~\bibnamefont {Zhu}}, \bibinfo {author} {\bibfnamefont {Y.}~\bibnamefont
  {Zhang}}, \bibinfo {author} {\bibfnamefont {K.}~\bibnamefont {Watanabe}},
  \bibinfo {author} {\bibfnamefont {T.}~\bibnamefont {Taniguchi}}, \bibinfo
  {author} {\bibfnamefont {M.~E.}\ \bibnamefont {Huber}},\ and\ \bibinfo
  {author} {\bibfnamefont {A.~F.}\ \bibnamefont {Young}},\ }\bibfield  {title}
  {\bibinfo {title} {Imaging orbital ferromagnetism in a moiré chern
  insulator},\ }\href {https://doi.org/10.1126/science.abd3190} {\bibfield
  {journal} {\bibinfo  {journal} {Science}\ }\textbf {\bibinfo {volume}
  {372}},\ \bibinfo {pages} {1323} (\bibinfo {year} {2021})}\BibitemShut
  {NoStop}%
\bibitem [{\citenamefont {Mei}\ \emph {et~al.}(2024)\citenamefont {Mei},
  \citenamefont {Zhao}, \citenamefont {Wang}, \citenamefont {Ren},
  \citenamefont {Xiao}, \citenamefont {Chang},\ and\ \citenamefont
  {Liu}}]{Mei2024}%
  \BibitemOpen
  \bibfield  {author} {\bibinfo {author} {\bibfnamefont {R.}~\bibnamefont
  {Mei}}, \bibinfo {author} {\bibfnamefont {Y.-F.}\ \bibnamefont {Zhao}},
  \bibinfo {author} {\bibfnamefont {C.}~\bibnamefont {Wang}}, \bibinfo {author}
  {\bibfnamefont {Y.}~\bibnamefont {Ren}}, \bibinfo {author} {\bibfnamefont
  {D.}~\bibnamefont {Xiao}}, \bibinfo {author} {\bibfnamefont {C.-Z.}\
  \bibnamefont {Chang}},\ and\ \bibinfo {author} {\bibfnamefont {C.-X.}\
  \bibnamefont {Liu}},\ }\bibfield  {title} {\bibinfo {title} {Electrically
  controlled anomalous hall effect and orbital magnetization in topological
  magnet $\mathrm{MnBi}_{2}\mathrm{Te}_{4}$},\ }\href
  {https://doi.org/10.1103/PhysRevLett.132.066604} {\bibfield  {journal}
  {\bibinfo  {journal} {Phys. Rev. Lett.}\ }\textbf {\bibinfo {volume} {132}},\
  \bibinfo {pages} {066604} (\bibinfo {year} {2024})}\BibitemShut {NoStop}%
\bibitem [{\citenamefont {Nakamura}\ \emph {et~al.}(2024)\citenamefont
  {Nakamura}, \citenamefont {Nishihaya}, \citenamefont {Ishizuka},
  \citenamefont {Kriener}, \citenamefont {Watanabe},\ and\ \citenamefont
  {Uchida}}]{Nakamura2024}%
  \BibitemOpen
  \bibfield  {author} {\bibinfo {author} {\bibfnamefont {A.}~\bibnamefont
  {Nakamura}}, \bibinfo {author} {\bibfnamefont {S.}~\bibnamefont {Nishihaya}},
  \bibinfo {author} {\bibfnamefont {H.}~\bibnamefont {Ishizuka}}, \bibinfo
  {author} {\bibfnamefont {M.}~\bibnamefont {Kriener}}, \bibinfo {author}
  {\bibfnamefont {Y.}~\bibnamefont {Watanabe}},\ and\ \bibinfo {author}
  {\bibfnamefont {M.}~\bibnamefont {Uchida}},\ }\bibfield  {title} {\bibinfo
  {title} {In-plane anomalous hall effect associated with orbital
  magnetization: Measurements of low-carrier density films of a magnetic weyl
  semimetal},\ }\href {https://doi.org/10.1103/PhysRevLett.133.236602}
  {\bibfield  {journal} {\bibinfo  {journal} {Phys. Rev. Lett.}\ }\textbf
  {\bibinfo {volume} {133}},\ \bibinfo {pages} {236602} (\bibinfo {year}
  {2024})}\BibitemShut {NoStop}%
\bibitem [{\citenamefont {Ye}\ \emph {et~al.}(2025)\citenamefont {Ye},
  \citenamefont {Tenzin}, \citenamefont {Sławińska},\ and\ \citenamefont
  {Autieri}}]{Ye2025}%
  \BibitemOpen
  \bibfield  {author} {\bibinfo {author} {\bibfnamefont {C.~C.}\ \bibnamefont
  {Ye}}, \bibinfo {author} {\bibfnamefont {K.}~\bibnamefont {Tenzin}}, \bibinfo
  {author} {\bibfnamefont {J.}~\bibnamefont {Sławińska}},\ and\ \bibinfo
  {author} {\bibfnamefont {C.}~\bibnamefont {Autieri}},\ }\bibfield  {title}
  {\bibinfo {title} {Dominant orbital magnetization in the prototypical
  altermagnet mnte},\ }\href {https://arxiv.org/abs/2505.08675} {\bibfield
  {journal} {\bibinfo  {journal} {arXiv:2505.08675 [cond-mat.mtrl-sci]}\ }
  (\bibinfo {year} {2025})}\BibitemShut {NoStop}%
\bibitem [{\citenamefont {Sorn}\ and\ \citenamefont
  {Mokrousov}(2025)}]{Sopheak2025}%
  \BibitemOpen
  \bibfield  {author} {\bibinfo {author} {\bibfnamefont {S.}~\bibnamefont
  {Sorn}}\ and\ \bibinfo {author} {\bibfnamefont {Y.}~\bibnamefont
  {Mokrousov}},\ }\bibfield  {title} {\bibinfo {title} {Activation of anomalous
  hall effect and orbital magnetization by domain walls in altermagnets},\
  }\href {https://arxiv.org/abs/2505.15894} {\bibfield  {journal} {\bibinfo
  {journal} {arXiv:2505.15894 [cond-mat.mes-hall]}\ } (\bibinfo {year}
  {2025})}\BibitemShut {NoStop}%
\bibitem [{\citenamefont {Bangar}\ \emph {et~al.}(2025)\citenamefont {Bangar},
  \citenamefont {Tsipas}, \citenamefont {Rout}, \citenamefont {Pandey},
  \citenamefont {Kalaboukhov}, \citenamefont {Lintzeris}, \citenamefont
  {Dimoulas},\ and\ \citenamefont {Dash}}]{Bangar2025}%
  \BibitemOpen
  \bibfield  {author} {\bibinfo {author} {\bibfnamefont {H.}~\bibnamefont
  {Bangar}}, \bibinfo {author} {\bibfnamefont {P.}~\bibnamefont {Tsipas}},
  \bibinfo {author} {\bibfnamefont {P.}~\bibnamefont {Rout}}, \bibinfo {author}
  {\bibfnamefont {L.}~\bibnamefont {Pandey}}, \bibinfo {author} {\bibfnamefont
  {A.}~\bibnamefont {Kalaboukhov}}, \bibinfo {author} {\bibfnamefont
  {A.}~\bibnamefont {Lintzeris}}, \bibinfo {author} {\bibfnamefont
  {A.}~\bibnamefont {Dimoulas}},\ and\ \bibinfo {author} {\bibfnamefont
  {S.~P.}\ \bibnamefont {Dash}},\ }\bibfield  {title} {\bibinfo {title}
  {Interplay between altermagnetic order and crystal symmetry probed using
  magnetotransport in epitaxial altermagnet mnte},\ }\href
  {https://arxiv.org/abs/2505.14589} {\bibfield  {journal} {\bibinfo  {journal}
  {arXiv:2505.14589 [cond-mat.mtrl-sci]}\ } (\bibinfo {year}
  {2025})}\BibitemShut {NoStop}%
\bibitem [{\citenamefont {Liu}\ \emph {et~al.}(2025{\natexlab{a}})\citenamefont
  {Liu}, \citenamefont {Wang}, \citenamefont {Zhang}, \citenamefont {Cao},\
  and\ \citenamefont {Xiao}}]{liu2025c}%
  \BibitemOpen
  \bibfield  {author} {\bibinfo {author} {\bibfnamefont {X.}~\bibnamefont
  {Liu}}, \bibinfo {author} {\bibfnamefont {C.}~\bibnamefont {Wang}}, \bibinfo
  {author} {\bibfnamefont {X.-W.}\ \bibnamefont {Zhang}}, \bibinfo {author}
  {\bibfnamefont {T.}~\bibnamefont {Cao}},\ and\ \bibinfo {author}
  {\bibfnamefont {D.}~\bibnamefont {Xiao}},\ }\bibfield  {title} {\bibinfo
  {title} {Orbital magnetization in correlated states of twisted bilayer
  transition metal dichalcogenides},\ }\href {https://arxiv.org/abs/2510.01727}
  {\bibfield  {journal} {\bibinfo  {journal} {arXiv:2510.01727
  [cond-mat.mes-hall]}\ } (\bibinfo {year} {2025}{\natexlab{a}})}\BibitemShut
  {NoStop}%
\bibitem [{\citenamefont {Zhu}\ and\ \citenamefont {Huang}(2025)}]{Zhu2025}%
  \BibitemOpen
  \bibfield  {author} {\bibinfo {author} {\bibfnamefont {J.}~\bibnamefont
  {Zhu}}\ and\ \bibinfo {author} {\bibfnamefont {C.}~\bibnamefont {Huang}},\
  }\bibfield  {title} {\bibinfo {title} {Magnetic-field-induced geometric
  response of mean-field projectors: Streda formula and orbital
  magnetization},\ }\href {https://arxiv.org/abs/2510.07001} {\bibfield
  {journal} {\bibinfo  {journal} {arXiv:2510.07001 [cond-mat.mes-hall]}\ }
  (\bibinfo {year} {2025})}\BibitemShut {NoStop}%
\bibitem [{\citenamefont {Kang}\ \emph {et~al.}(2025)\citenamefont {Kang},
  \citenamefont {Wang},\ and\ \citenamefont {Vafek}}]{Kang2025}%
  \BibitemOpen
  \bibfield  {author} {\bibinfo {author} {\bibfnamefont {J.}~\bibnamefont
  {Kang}}, \bibinfo {author} {\bibfnamefont {M.}~\bibnamefont {Wang}},\ and\
  \bibinfo {author} {\bibfnamefont {O.}~\bibnamefont {Vafek}},\ }\bibfield
  {title} {\bibinfo {title} {Orbital magnetization and magnetic susceptibility
  of interacting electrons},\ }\href {https://arxiv.org/abs/2509.20626}
  {\bibfield  {journal} {\bibinfo  {journal} {arXiv:2509.20626
  [cond-mat.str-el]}\ } (\bibinfo {year} {2025})}\BibitemShut {NoStop}%
\bibitem [{\citenamefont {Lage}\ \emph {et~al.}(2025)\citenamefont {Lage},
  \citenamefont {Cysne},\ and\ \citenamefont {Latgé}}]{Lage2025}%
  \BibitemOpen
  \bibfield  {author} {\bibinfo {author} {\bibfnamefont {L.~L.}\ \bibnamefont
  {Lage}}, \bibinfo {author} {\bibfnamefont {T.~P.}\ \bibnamefont {Cysne}},\
  and\ \bibinfo {author} {\bibfnamefont {A.}~\bibnamefont {Latgé}},\
  }\bibfield  {title} {\bibinfo {title} {Orbital magnetization in sierpinski
  fractals},\ }\href {https://arxiv.org/abs/2510.14556} {\bibfield  {journal}
  {\bibinfo  {journal} {arXiv:2510.14556 [cond-mat.mtrl-sci]}\ } (\bibinfo
  {year} {2025})}\BibitemShut {NoStop}%
\bibitem [{\citenamefont {Lu}\ \emph {et~al.}(2025)\citenamefont {Lu},
  \citenamefont {Jiang}, \citenamefont {Guo},\ and\ \citenamefont
  {Liu}}]{Lu2025}%
  \BibitemOpen
  \bibfield  {author} {\bibinfo {author} {\bibfnamefont {X.}~\bibnamefont
  {Lu}}, \bibinfo {author} {\bibfnamefont {R.}~\bibnamefont {Jiang}}, \bibinfo
  {author} {\bibfnamefont {Z.}~\bibnamefont {Guo}},\ and\ \bibinfo {author}
  {\bibfnamefont {J.}~\bibnamefont {Liu}},\ }\bibfield  {title} {\bibinfo
  {title} {Orbital magnetoelectric coupling of three dimensional chern
  insulators},\ }\href {https://doi.org/10.1038/s41535-025-00794-z} {\bibfield
  {journal} {\bibinfo  {journal} {npj Quantum Materials}\ }\textbf {\bibinfo
  {volume} {10}},\ \bibinfo {pages} {76} (\bibinfo {year} {2025})}\BibitemShut
  {NoStop}%
\bibitem [{\citenamefont {Malashevich}\ \emph {et~al.}(2010)\citenamefont
  {Malashevich}, \citenamefont {Souza}, \citenamefont {Coh},\ and\
  \citenamefont {Vanderbilt}}]{Malashevich2010}%
  \BibitemOpen
  \bibfield  {author} {\bibinfo {author} {\bibfnamefont {A.}~\bibnamefont
  {Malashevich}}, \bibinfo {author} {\bibfnamefont {I.}~\bibnamefont {Souza}},
  \bibinfo {author} {\bibfnamefont {S.}~\bibnamefont {Coh}},\ and\ \bibinfo
  {author} {\bibfnamefont {D.}~\bibnamefont {Vanderbilt}},\ }\bibfield  {title}
  {\bibinfo {title} {Theory of orbital magnetoelectric response},\ }\href
  {https://doi.org/10.1088/1367-2630/12/5/053032} {\bibfield  {journal}
  {\bibinfo  {journal} {New Journal of Physics}\ }\textbf {\bibinfo {volume}
  {12}},\ \bibinfo {pages} {053032} (\bibinfo {year} {2010})}\BibitemShut
  {NoStop}%
\bibitem [{\citenamefont {Bhowal}\ and\ \citenamefont
  {Satpathy}(2020)}]{Bhowal2020}%
  \BibitemOpen
  \bibfield  {author} {\bibinfo {author} {\bibfnamefont {S.}~\bibnamefont
  {Bhowal}}\ and\ \bibinfo {author} {\bibfnamefont {S.}~\bibnamefont
  {Satpathy}},\ }\bibfield  {title} {\bibinfo {title} {Intrinsic orbital and
  spin hall effects in monolayer transition metal dichalcogenides},\ }\href
  {https://doi.org/10.1103/PhysRevB.102.035409} {\bibfield  {journal} {\bibinfo
   {journal} {Phys. Rev. B}\ }\textbf {\bibinfo {volume} {102}},\ \bibinfo
  {pages} {035409} (\bibinfo {year} {2020})}\BibitemShut {NoStop}%
\bibitem [{\citenamefont {Cysne}\ \emph {et~al.}(2022)\citenamefont {Cysne},
  \citenamefont {Bhowal}, \citenamefont {Vignale},\ and\ \citenamefont
  {Rappoport}}]{Cysne2022}%
  \BibitemOpen
  \bibfield  {author} {\bibinfo {author} {\bibfnamefont {T.~P.}\ \bibnamefont
  {Cysne}}, \bibinfo {author} {\bibfnamefont {S.}~\bibnamefont {Bhowal}},
  \bibinfo {author} {\bibfnamefont {G.}~\bibnamefont {Vignale}},\ and\ \bibinfo
  {author} {\bibfnamefont {T.~G.}\ \bibnamefont {Rappoport}},\ }\bibfield
  {title} {\bibinfo {title} {Orbital hall effect in bilayer transition metal
  dichalcogenides: From the intra-atomic approximation to the bloch states
  orbital magnetic moment approach},\ }\href
  {https://doi.org/10.1103/PhysRevB.105.195421} {\bibfield  {journal} {\bibinfo
   {journal} {Phys. Rev. B}\ }\textbf {\bibinfo {volume} {105}},\ \bibinfo
  {pages} {195421} (\bibinfo {year} {2022})}\BibitemShut {NoStop}%
\bibitem [{\citenamefont {Pezo}\ \emph {et~al.}(2023)\citenamefont {Pezo},
  \citenamefont {Garc\'{\i}a~Ovalle},\ and\ \citenamefont
  {Manchon}}]{Pezo2023}%
  \BibitemOpen
  \bibfield  {author} {\bibinfo {author} {\bibfnamefont {A.}~\bibnamefont
  {Pezo}}, \bibinfo {author} {\bibfnamefont {D.}~\bibnamefont
  {Garc\'{\i}a~Ovalle}},\ and\ \bibinfo {author} {\bibfnamefont
  {A.}~\bibnamefont {Manchon}},\ }\bibfield  {title} {\bibinfo {title} {Orbital
  hall physics in two-dimensional dirac materials},\ }\href
  {https://doi.org/10.1103/PhysRevB.108.075427} {\bibfield  {journal} {\bibinfo
   {journal} {Phys. Rev. B}\ }\textbf {\bibinfo {volume} {108}},\ \bibinfo
  {pages} {075427} (\bibinfo {year} {2023})}\BibitemShut {NoStop}%
\bibitem [{\citenamefont {Busch}\ \emph {et~al.}(2023)\citenamefont {Busch},
  \citenamefont {Mertig},\ and\ \citenamefont {G\"obel}}]{Busch2023}%
  \BibitemOpen
  \bibfield  {author} {\bibinfo {author} {\bibfnamefont {O.}~\bibnamefont
  {Busch}}, \bibinfo {author} {\bibfnamefont {I.}~\bibnamefont {Mertig}},\ and\
  \bibinfo {author} {\bibfnamefont {B.}~\bibnamefont {G\"obel}},\ }\bibfield
  {title} {\bibinfo {title} {Orbital hall effect and orbital edge states caused
  by $s$ electrons},\ }\href {https://doi.org/10.1103/PhysRevResearch.5.043052}
  {\bibfield  {journal} {\bibinfo  {journal} {Phys. Rev. Res.}\ }\textbf
  {\bibinfo {volume} {5}},\ \bibinfo {pages} {043052} (\bibinfo {year}
  {2023})}\BibitemShut {NoStop}%
\bibitem [{\citenamefont {Liu}\ \emph {et~al.}(2025{\natexlab{b}})\citenamefont
  {Liu}, \citenamefont {Cullen}, \citenamefont {Arovas},\ and\ \citenamefont
  {Culcer}}]{Liu2025}%
  \BibitemOpen
  \bibfield  {author} {\bibinfo {author} {\bibfnamefont {H.}~\bibnamefont
  {Liu}}, \bibinfo {author} {\bibfnamefont {J.~H.}\ \bibnamefont {Cullen}},
  \bibinfo {author} {\bibfnamefont {D.~P.}\ \bibnamefont {Arovas}},\ and\
  \bibinfo {author} {\bibfnamefont {D.}~\bibnamefont {Culcer}},\ }\bibfield
  {title} {\bibinfo {title} {Quantum correction to the orbital hall effect},\
  }\href {https://doi.org/10.1103/PhysRevLett.134.036304} {\bibfield  {journal}
  {\bibinfo  {journal} {Phys. Rev. Lett.}\ }\textbf {\bibinfo {volume} {134}},\
  \bibinfo {pages} {036304} (\bibinfo {year} {2025}{\natexlab{b}})}\BibitemShut
  {NoStop}%
\bibitem [{\citenamefont {Cysne}\ \emph {et~al.}(2025)\citenamefont {Cysne},
  \citenamefont {Canonico}, \citenamefont {Costa}, \citenamefont {Muniz},\ and\
  \citenamefont {Rappoport}}]{Cysne2025}%
  \BibitemOpen
  \bibfield  {author} {\bibinfo {author} {\bibfnamefont {T.~P.}\ \bibnamefont
  {Cysne}}, \bibinfo {author} {\bibfnamefont {L.~M.}\ \bibnamefont {Canonico}},
  \bibinfo {author} {\bibfnamefont {M.}~\bibnamefont {Costa}}, \bibinfo
  {author} {\bibfnamefont {R.~B.}\ \bibnamefont {Muniz}},\ and\ \bibinfo
  {author} {\bibfnamefont {T.~G.}\ \bibnamefont {Rappoport}},\ }\bibfield
  {title} {\bibinfo {title} {Orbitronics in two-dimensional materials},\ }\href
  {https://doi.org/10.1038/s44306-025-00103-1} {\bibfield  {journal} {\bibinfo
  {journal} {npj Spintronics}\ }\textbf {\bibinfo {volume} {3}},\ \bibinfo
  {pages} {39} (\bibinfo {year} {2025})}\BibitemShut {NoStop}%
\bibitem [{\citenamefont {Veneri}\ \emph {et~al.}(2025)\citenamefont {Veneri},
  \citenamefont {Rappoport},\ and\ \citenamefont {Ferreira}}]{Veneri2025}%
  \BibitemOpen
  \bibfield  {author} {\bibinfo {author} {\bibfnamefont {A.}~\bibnamefont
  {Veneri}}, \bibinfo {author} {\bibfnamefont {T.~G.}\ \bibnamefont
  {Rappoport}},\ and\ \bibinfo {author} {\bibfnamefont {A.}~\bibnamefont
  {Ferreira}},\ }\bibfield  {title} {\bibinfo {title} {Extrinsic orbital hall
  effect: Orbital skew scattering and crossover between diffusive and intrinsic
  orbital transport},\ }\href {https://doi.org/10.1103/PhysRevLett.134.136201}
  {\bibfield  {journal} {\bibinfo  {journal} {Phys. Rev. Lett.}\ }\textbf
  {\bibinfo {volume} {134}},\ \bibinfo {pages} {136201} (\bibinfo {year}
  {2025})}\BibitemShut {NoStop}%
\bibitem [{\citenamefont {Xiao}\ \emph {et~al.}(2005)\citenamefont {Xiao},
  \citenamefont {Shi},\ and\ \citenamefont {Niu}}]{Xiao2005}%
  \BibitemOpen
  \bibfield  {author} {\bibinfo {author} {\bibfnamefont {D.}~\bibnamefont
  {Xiao}}, \bibinfo {author} {\bibfnamefont {J.}~\bibnamefont {Shi}},\ and\
  \bibinfo {author} {\bibfnamefont {Q.}~\bibnamefont {Niu}},\ }\bibfield
  {title} {\bibinfo {title} {Berry phase correction to electron density of
  states in solids},\ }\href {https://doi.org/10.1103/PhysRevLett.95.137204}
  {\bibfield  {journal} {\bibinfo  {journal} {Phys. Rev. Lett.}\ }\textbf
  {\bibinfo {volume} {95}},\ \bibinfo {pages} {137204} (\bibinfo {year}
  {2005})}\BibitemShut {NoStop}%
\bibitem [{\citenamefont {Thonhauser}\ \emph {et~al.}(2005)\citenamefont
  {Thonhauser}, \citenamefont {Ceresoli}, \citenamefont {Vanderbilt},\ and\
  \citenamefont {Resta}}]{Thonhauser2005}%
  \BibitemOpen
  \bibfield  {author} {\bibinfo {author} {\bibfnamefont {T.}~\bibnamefont
  {Thonhauser}}, \bibinfo {author} {\bibfnamefont {D.}~\bibnamefont
  {Ceresoli}}, \bibinfo {author} {\bibfnamefont {D.}~\bibnamefont
  {Vanderbilt}},\ and\ \bibinfo {author} {\bibfnamefont {R.}~\bibnamefont
  {Resta}},\ }\bibfield  {title} {\bibinfo {title} {Orbital magnetization in
  periodic insulators},\ }\href {https://doi.org/10.1103/PhysRevLett.95.137205}
  {\bibfield  {journal} {\bibinfo  {journal} {Phys. Rev. Lett.}\ }\textbf
  {\bibinfo {volume} {95}},\ \bibinfo {pages} {137205} (\bibinfo {year}
  {2005})}\BibitemShut {NoStop}%
\bibitem [{\citenamefont {Shi}\ \emph {et~al.}(2007)\citenamefont {Shi},
  \citenamefont {Vignale}, \citenamefont {Xiao},\ and\ \citenamefont
  {Niu}}]{Shi2007}%
  \BibitemOpen
  \bibfield  {author} {\bibinfo {author} {\bibfnamefont {J.}~\bibnamefont
  {Shi}}, \bibinfo {author} {\bibfnamefont {G.}~\bibnamefont {Vignale}},
  \bibinfo {author} {\bibfnamefont {D.}~\bibnamefont {Xiao}},\ and\ \bibinfo
  {author} {\bibfnamefont {Q.}~\bibnamefont {Niu}},\ }\bibfield  {title}
  {\bibinfo {title} {Quantum theory of orbital magnetization and its
  generalization to interacting systems},\ }\href
  {https://doi.org/10.1103/PhysRevLett.99.197202} {\bibfield  {journal}
  {\bibinfo  {journal} {Phys. Rev. Lett.}\ }\textbf {\bibinfo {volume} {99}},\
  \bibinfo {pages} {197202} (\bibinfo {year} {2007})}\BibitemShut {NoStop}%
\bibitem [{\citenamefont {Lopez}\ \emph {et~al.}(2012)\citenamefont {Lopez},
  \citenamefont {Vanderbilt}, \citenamefont {Thonhauser},\ and\ \citenamefont
  {Souza}}]{Lopez2012}%
  \BibitemOpen
  \bibfield  {author} {\bibinfo {author} {\bibfnamefont {M.~G.}\ \bibnamefont
  {Lopez}}, \bibinfo {author} {\bibfnamefont {D.}~\bibnamefont {Vanderbilt}},
  \bibinfo {author} {\bibfnamefont {T.}~\bibnamefont {Thonhauser}},\ and\
  \bibinfo {author} {\bibfnamefont {I.}~\bibnamefont {Souza}},\ }\bibfield
  {title} {\bibinfo {title} {Wannier-based calculation of the orbital
  magnetization in crystals},\ }\href
  {https://doi.org/10.1103/PhysRevB.85.014435} {\bibfield  {journal} {\bibinfo
  {journal} {Phys. Rev. B}\ }\textbf {\bibinfo {volume} {85}},\ \bibinfo
  {pages} {014435} (\bibinfo {year} {2012})}\BibitemShut {NoStop}%
\bibitem [{\citenamefont {Hanke}\ \emph {et~al.}(2016)\citenamefont {Hanke},
  \citenamefont {Freimuth}, \citenamefont {Nandy}, \citenamefont {Zhang},
  \citenamefont {Bl\"ugel},\ and\ \citenamefont {Mokrousov}}]{Hanke2016}%
  \BibitemOpen
  \bibfield  {author} {\bibinfo {author} {\bibfnamefont {J.-P.}\ \bibnamefont
  {Hanke}}, \bibinfo {author} {\bibfnamefont {F.}~\bibnamefont {Freimuth}},
  \bibinfo {author} {\bibfnamefont {A.~K.}\ \bibnamefont {Nandy}}, \bibinfo
  {author} {\bibfnamefont {H.}~\bibnamefont {Zhang}}, \bibinfo {author}
  {\bibfnamefont {S.}~\bibnamefont {Bl\"ugel}},\ and\ \bibinfo {author}
  {\bibfnamefont {Y.}~\bibnamefont {Mokrousov}},\ }\bibfield  {title} {\bibinfo
  {title} {Role of berry phase theory for describing orbital magnetism: From
  magnetic heterostructures to topological orbital ferromagnets},\ }\href
  {https://doi.org/10.1103/PhysRevB.94.121114} {\bibfield  {journal} {\bibinfo
  {journal} {Phys. Rev. B}\ }\textbf {\bibinfo {volume} {94}},\ \bibinfo
  {pages} {121114} (\bibinfo {year} {2016})}\BibitemShut {NoStop}%
\bibitem [{\citenamefont {Bianco}\ and\ \citenamefont
  {Resta}(2013)}]{Bianco2013}%
  \BibitemOpen
  \bibfield  {author} {\bibinfo {author} {\bibfnamefont {R.}~\bibnamefont
  {Bianco}}\ and\ \bibinfo {author} {\bibfnamefont {R.}~\bibnamefont {Resta}},\
  }\bibfield  {title} {\bibinfo {title} {Orbital magnetization as a local
  property},\ }\href {https://doi.org/10.1103/PhysRevLett.110.087202}
  {\bibfield  {journal} {\bibinfo  {journal} {Phys. Rev. Lett.}\ }\textbf
  {\bibinfo {volume} {110}},\ \bibinfo {pages} {087202} (\bibinfo {year}
  {2013})}\BibitemShut {NoStop}%
\bibitem [{\citenamefont {Wang}\ \emph {et~al.}(2022)\citenamefont {Wang},
  \citenamefont {Yu}, \citenamefont {Guan}, \citenamefont {Dai}, \citenamefont
  {Wang},\ and\ \citenamefont {Zhang}}]{Wang2022}%
  \BibitemOpen
  \bibfield  {author} {\bibinfo {author} {\bibfnamefont {S.-S.}\ \bibnamefont
  {Wang}}, \bibinfo {author} {\bibfnamefont {Y.}~\bibnamefont {Yu}}, \bibinfo
  {author} {\bibfnamefont {J.-H.}\ \bibnamefont {Guan}}, \bibinfo {author}
  {\bibfnamefont {Y.-M.}\ \bibnamefont {Dai}}, \bibinfo {author} {\bibfnamefont
  {H.-H.}\ \bibnamefont {Wang}},\ and\ \bibinfo {author} {\bibfnamefont
  {Y.-Y.}\ \bibnamefont {Zhang}},\ }\bibfield  {title} {\bibinfo {title}
  {Boundary effects on orbital magnetization for a bilayer system with
  different chern numbers},\ }\href
  {https://doi.org/10.1103/PhysRevB.106.075136} {\bibfield  {journal} {\bibinfo
   {journal} {Phys. Rev. B}\ }\textbf {\bibinfo {volume} {106}},\ \bibinfo
  {pages} {075136} (\bibinfo {year} {2022})}\BibitemShut {NoStop}%
\bibitem [{\citenamefont {Marrazzo}\ and\ \citenamefont
  {Resta}(2016)}]{Marrazzo2016}%
  \BibitemOpen
  \bibfield  {author} {\bibinfo {author} {\bibfnamefont {A.}~\bibnamefont
  {Marrazzo}}\ and\ \bibinfo {author} {\bibfnamefont {R.}~\bibnamefont
  {Resta}},\ }\bibfield  {title} {\bibinfo {title} {Irrelevance of the boundary
  on the magnetization of metals},\ }\href
  {https://doi.org/10.1103/PhysRevLett.116.137201} {\bibfield  {journal}
  {\bibinfo  {journal} {Phys. Rev. Lett.}\ }\textbf {\bibinfo {volume} {116}},\
  \bibinfo {pages} {137201} (\bibinfo {year} {2016})}\BibitemShut {NoStop}%
\bibitem [{\citenamefont {Wang}\ \emph {et~al.}(2023)\citenamefont {Wang},
  \citenamefont {Dai}, \citenamefont {Wang}, \citenamefont {Chen},
  \citenamefont {Zhang},\ and\ \citenamefont {Zhang}}]{Wang2023}%
  \BibitemOpen
  \bibfield  {author} {\bibinfo {author} {\bibfnamefont {S.-S.}\ \bibnamefont
  {Wang}}, \bibinfo {author} {\bibfnamefont {Y.-M.}\ \bibnamefont {Dai}},
  \bibinfo {author} {\bibfnamefont {H.-H.}\ \bibnamefont {Wang}}, \bibinfo
  {author} {\bibfnamefont {H.-C.}\ \bibnamefont {Chen}}, \bibinfo {author}
  {\bibfnamefont {B.}~\bibnamefont {Zhang}},\ and\ \bibinfo {author}
  {\bibfnamefont {Y.-Y.}\ \bibnamefont {Zhang}},\ }\bibfield  {title} {\bibinfo
  {title} {Orbital magnetization under electric field and orbital
  magnetoelectric polarizability for a bilayer chern system},\ }\href
  {https://doi.org/10.1103/PhysRevB.107.125135} {\bibfield  {journal} {\bibinfo
   {journal} {Phys. Rev. B}\ }\textbf {\bibinfo {volume} {107}},\ \bibinfo
  {pages} {125135} (\bibinfo {year} {2023})}\BibitemShut {NoStop}%
\bibitem [{\citenamefont {Ceresoli}\ and\ \citenamefont
  {Resta}(2007)}]{Ceresoli2007}%
  \BibitemOpen
  \bibfield  {author} {\bibinfo {author} {\bibfnamefont {D.}~\bibnamefont
  {Ceresoli}}\ and\ \bibinfo {author} {\bibfnamefont {R.}~\bibnamefont
  {Resta}},\ }\bibfield  {title} {\bibinfo {title} {Orbital magnetization and
  chern number in a supercell framework: Single $\mathbf{k}$-point formula},\
  }\href {https://doi.org/10.1103/PhysRevB.76.012405} {\bibfield  {journal}
  {\bibinfo  {journal} {Phys. Rev. B}\ }\textbf {\bibinfo {volume} {76}},\
  \bibinfo {pages} {012405} (\bibinfo {year} {2007})}\BibitemShut {NoStop}%
\bibitem [{\citenamefont {Seleznev}\ and\ \citenamefont
  {Vanderbilt}(2023)}]{Seleznev2023}%
  \BibitemOpen
  \bibfield  {author} {\bibinfo {author} {\bibfnamefont {D.}~\bibnamefont
  {Seleznev}}\ and\ \bibinfo {author} {\bibfnamefont {D.}~\bibnamefont
  {Vanderbilt}},\ }\bibfield  {title} {\bibinfo {title} {Towards a theory of
  surface orbital magnetization},\ }\href
  {https://doi.org/10.1103/PhysRevB.107.115102} {\bibfield  {journal} {\bibinfo
   {journal} {Phys. Rev. B}\ }\textbf {\bibinfo {volume} {107}},\ \bibinfo
  {pages} {115102} (\bibinfo {year} {2023})}\BibitemShut {NoStop}%
\bibitem [{\citenamefont {Weiße}\ \emph {et~al.}(2006)\citenamefont {Weiße},
  \citenamefont {Wellein}, \citenamefont {Alvermann},\ and\ \citenamefont
  {Fehske}}]{Weise2006}%
  \BibitemOpen
  \bibfield  {author} {\bibinfo {author} {\bibfnamefont {A.}~\bibnamefont
  {Weiße}}, \bibinfo {author} {\bibfnamefont {G.}~\bibnamefont {Wellein}},
  \bibinfo {author} {\bibfnamefont {A.}~\bibnamefont {Alvermann}},\ and\
  \bibinfo {author} {\bibfnamefont {H.}~\bibnamefont {Fehske}},\ }\bibfield
  {title} {\bibinfo {title} {The kernel polynomial method},\ }\href
  {https://doi.org/10.1103/RevModPhys.78.275} {\bibfield  {journal} {\bibinfo
  {journal} {Reviews of Modern Physics}\ }\textbf {\bibinfo {volume} {78}},\
  \bibinfo {pages} {275} (\bibinfo {year} {2006})},\ \bibinfo {note}
  {publisher: American Physical Society}\BibitemShut {NoStop}%
\bibitem [{\citenamefont {García}\ \emph {et~al.}(2015)\citenamefont
  {García}, \citenamefont {Covaci},\ and\ \citenamefont
  {Rappoport}}]{Garcia2015}%
  \BibitemOpen
  \bibfield  {author} {\bibinfo {author} {\bibfnamefont {J.~H.}\ \bibnamefont
  {García}}, \bibinfo {author} {\bibfnamefont {L.}~\bibnamefont {Covaci}},\
  and\ \bibinfo {author} {\bibfnamefont {T.~G.}\ \bibnamefont {Rappoport}},\
  }\bibfield  {title} {\bibinfo {title} {Real-{Space} {Calculation} of the
  {Conductivity} {Tensor} for {Disordered} {Topological} {Matter}},\ }\href
  {https://doi.org/10.1103/PhysRevLett.114.116602} {\bibfield  {journal}
  {\bibinfo  {journal} {Physical Review Letters}\ }\textbf {\bibinfo {volume}
  {114}},\ \bibinfo {pages} {116602} (\bibinfo {year} {2015})},\ \bibinfo
  {note} {publisher: American Physical Society}\BibitemShut {NoStop}%
\bibitem [{\citenamefont {Ferreira}\ and\ \citenamefont
  {Mucciolo}(2015)}]{Ferreira2015}%
  \BibitemOpen
  \bibfield  {author} {\bibinfo {author} {\bibfnamefont {A.}~\bibnamefont
  {Ferreira}}\ and\ \bibinfo {author} {\bibfnamefont {E.~R.}\ \bibnamefont
  {Mucciolo}},\ }\bibfield  {title} {\bibinfo {title} {Critical delocalization
  of chiral zero energy modes in graphene},\ }\href
  {https://doi.org/10.1103/PhysRevLett.115.106601} {\bibfield  {journal}
  {\bibinfo  {journal} {Phys. Rev. Lett.}\ }\textbf {\bibinfo {volume} {115}},\
  \bibinfo {pages} {106601} (\bibinfo {year} {2015})}\BibitemShut {NoStop}%
\bibitem [{\citenamefont {Jo{\~{a}}o}\ \emph {et~al.}(2020)\citenamefont
  {Jo{\~{a}}o}, \citenamefont {An{\dj}elkovi{\'{c}}}, \citenamefont {Covaci},
  \citenamefont {Rappoport}, \citenamefont {Lopes},\ and\ \citenamefont
  {Ferreira}}]{kite}%
  \BibitemOpen
  \bibfield  {author} {\bibinfo {author} {\bibfnamefont {S.~M.}\ \bibnamefont
  {Jo{\~{a}}o}}, \bibinfo {author} {\bibfnamefont {M.}~\bibnamefont
  {An{\dj}elkovi{\'{c}}}}, \bibinfo {author} {\bibfnamefont {L.}~\bibnamefont
  {Covaci}}, \bibinfo {author} {\bibfnamefont {T.~G.}\ \bibnamefont
  {Rappoport}}, \bibinfo {author} {\bibfnamefont {J.~M. V.~P.}\ \bibnamefont
  {Lopes}},\ and\ \bibinfo {author} {\bibfnamefont {A.}~\bibnamefont
  {Ferreira}},\ }\bibfield  {title} {\bibinfo {title} {{KITE}: high-performance
  accurate modelling of electronic structure and response functions of large
  molecules, disordered crystals and heterostructures},\ }\href
  {https://doi.org/10.1098/rsos.191809} {\bibfield  {journal} {\bibinfo
  {journal} {Royal Society Open Science}\ }\textbf {\bibinfo {volume} {7}},\
  \bibinfo {pages} {191809} (\bibinfo {year} {2020})}\BibitemShut {NoStop}%
\bibitem [{\citenamefont {Fan}\ \emph {et~al.}(2021)\citenamefont {Fan},
  \citenamefont {Garcia}, \citenamefont {Cummings}, \citenamefont
  {Barrios-Vargas}, \citenamefont {Panhans}, \citenamefont {Harju},
  \citenamefont {Ortmann},\ and\ \citenamefont {Roche}}]{fan2021linear}%
  \BibitemOpen
  \bibfield  {author} {\bibinfo {author} {\bibfnamefont {Z.}~\bibnamefont
  {Fan}}, \bibinfo {author} {\bibfnamefont {J.~H.}\ \bibnamefont {Garcia}},
  \bibinfo {author} {\bibfnamefont {A.~W.}\ \bibnamefont {Cummings}}, \bibinfo
  {author} {\bibfnamefont {J.~E.}\ \bibnamefont {Barrios-Vargas}}, \bibinfo
  {author} {\bibfnamefont {M.}~\bibnamefont {Panhans}}, \bibinfo {author}
  {\bibfnamefont {A.}~\bibnamefont {Harju}}, \bibinfo {author} {\bibfnamefont
  {F.}~\bibnamefont {Ortmann}},\ and\ \bibinfo {author} {\bibfnamefont
  {S.}~\bibnamefont {Roche}},\ }\bibfield  {title} {\bibinfo {title} {Linear
  scaling quantum transport methodologies},\ }\href
  {https://doi.org/https://doi.org/10.1016/j.physrep.2020.12.001} {\bibfield
  {journal} {\bibinfo  {journal} {Physics Reports}\ }\textbf {\bibinfo {volume}
  {903}},\ \bibinfo {pages} {1} (\bibinfo {year} {2021})}\BibitemShut {NoStop}%
\bibitem [{\citenamefont {Santos~Pires}\ \emph {et~al.}(2022)\citenamefont
  {Santos~Pires}, \citenamefont {Jo\~ao}, \citenamefont {Ferreira},
  \citenamefont {Amorim},\ and\ \citenamefont {Viana
  Parente~Lopes}}]{Pires2022}%
  \BibitemOpen
  \bibfield  {author} {\bibinfo {author} {\bibfnamefont {J.~P.}\ \bibnamefont
  {Santos~Pires}}, \bibinfo {author} {\bibfnamefont {S.~M.}\ \bibnamefont
  {Jo\~ao}}, \bibinfo {author} {\bibfnamefont {A.}~\bibnamefont {Ferreira}},
  \bibinfo {author} {\bibfnamefont {B.}~\bibnamefont {Amorim}},\ and\ \bibinfo
  {author} {\bibfnamefont {J.~M.}\ \bibnamefont {Viana Parente~Lopes}},\
  }\bibfield  {title} {\bibinfo {title} {Anomalous transport signatures in weyl
  semimetals with point defects},\ }\href
  {https://doi.org/10.1103/PhysRevLett.129.196601} {\bibfield  {journal}
  {\bibinfo  {journal} {Phys. Rev. Lett.}\ }\textbf {\bibinfo {volume} {129}},\
  \bibinfo {pages} {196601} (\bibinfo {year} {2022})}\BibitemShut {NoStop}%
\bibitem [{\citenamefont {de~Castro}\ \emph {et~al.}(2024)\citenamefont
  {de~Castro}, \citenamefont {Lopes}, \citenamefont {Ferreira},\ and\
  \citenamefont {Bahamon}}]{Castro2024}%
  \BibitemOpen
  \bibfield  {author} {\bibinfo {author} {\bibfnamefont {S.~G.}\ \bibnamefont
  {de~Castro}}, \bibinfo {author} {\bibfnamefont {J.~a. M. V.~P.}\ \bibnamefont
  {Lopes}}, \bibinfo {author} {\bibfnamefont {A.}~\bibnamefont {Ferreira}},\
  and\ \bibinfo {author} {\bibfnamefont {D.~A.}\ \bibnamefont {Bahamon}},\
  }\bibfield  {title} {\bibinfo {title} {Fast fourier-chebyshev approach to
  real-space simulations of the kubo formula},\ }\href
  {https://doi.org/10.1103/PhysRevLett.132.076302} {\bibfield  {journal}
  {\bibinfo  {journal} {Phys. Rev. Lett.}\ }\textbf {\bibinfo {volume} {132}},\
  \bibinfo {pages} {076302} (\bibinfo {year} {2024})}\BibitemShut {NoStop}%
\bibitem [{\citenamefont {Canonico}\ \emph {et~al.}(2024)\citenamefont
  {Canonico}, \citenamefont {Garcia},\ and\ \citenamefont
  {Roche}}]{Canonico2024}%
  \BibitemOpen
  \bibfield  {author} {\bibinfo {author} {\bibfnamefont {L.~M.}\ \bibnamefont
  {Canonico}}, \bibinfo {author} {\bibfnamefont {J.~H.}\ \bibnamefont
  {Garcia}},\ and\ \bibinfo {author} {\bibfnamefont {S.}~\bibnamefont
  {Roche}},\ }\bibfield  {title} {\bibinfo {title} {Orbital hall responses in
  disordered topological materials},\ }\href
  {https://doi.org/10.1103/PhysRevB.110.L140201} {\bibfield  {journal}
  {\bibinfo  {journal} {Phys. Rev. B}\ }\textbf {\bibinfo {volume} {110}},\
  \bibinfo {pages} {L140201} (\bibinfo {year} {2024})}\BibitemShut {NoStop}%
\bibitem [{\citenamefont {Cardias}\ \emph {et~al.}(2025)\citenamefont
  {Cardias}, \citenamefont {Strand}, \citenamefont {Bergman}, \citenamefont
  {Klautau},\ and\ \citenamefont {Rappoport}}]{Cardias2025}%
  \BibitemOpen
  \bibfield  {author} {\bibinfo {author} {\bibfnamefont {R.}~\bibnamefont
  {Cardias}}, \bibinfo {author} {\bibfnamefont {H.~U.~R.}\ \bibnamefont
  {Strand}}, \bibinfo {author} {\bibfnamefont {A.}~\bibnamefont {Bergman}},
  \bibinfo {author} {\bibfnamefont {A.~B.}\ \bibnamefont {Klautau}},\ and\
  \bibinfo {author} {\bibfnamefont {T.~G.}\ \bibnamefont {Rappoport}},\
  }\bibfield  {title} {\bibinfo {title} {Real-space first-principles approach
  to orbitronic phenomena in metallic multilayers},\ }\href
  {https://arxiv.org/abs/2508.14270} {\bibfield  {journal} {\bibinfo  {journal}
  {arXiv:2508.14270 [cond-mat.mtrl-sci]}\ } (\bibinfo {year}
  {2025})}\BibitemShut {NoStop}%
\bibitem [{\citenamefont {Resta}(2010)}]{Resta_2010}%
  \BibitemOpen
  \bibfield  {author} {\bibinfo {author} {\bibfnamefont {R.}~\bibnamefont
  {Resta}},\ }\bibfield  {title} {\bibinfo {title} {Electrical polarization and
  orbital magnetization: the modern theories},\ }\href
  {https://doi.org/10.1088/0953-8984/22/12/123201} {\bibfield  {journal}
  {\bibinfo  {journal} {Journal of Physics: Condensed Matter}\ }\textbf
  {\bibinfo {volume} {22}},\ \bibinfo {pages} {123201} (\bibinfo {year}
  {2010})}\BibitemShut {NoStop}%
\bibitem [{\citenamefont {Haldane}(1988)}]{Haldane1988}%
  \BibitemOpen
  \bibfield  {author} {\bibinfo {author} {\bibfnamefont {F.~D.~M.}\
  \bibnamefont {Haldane}},\ }\bibfield  {title} {\bibinfo {title} {Model for a
  quantum hall effect without landau levels: Condensed-matter realization of
  the "parity anomaly"},\ }\href {https://doi.org/10.1103/PhysRevLett.61.2015}
  {\bibfield  {journal} {\bibinfo  {journal} {Phys. Rev. Lett.}\ }\textbf
  {\bibinfo {volume} {61}},\ \bibinfo {pages} {2015} (\bibinfo {year}
  {1988})}\BibitemShut {NoStop}%
\bibitem [{\citenamefont {Kitaev}(2006)}]{Kitaev2006}%
  \BibitemOpen
  \bibfield  {author} {\bibinfo {author} {\bibfnamefont {A.}~\bibnamefont
  {Kitaev}},\ }\bibfield  {title} {\bibinfo {title} {Anyons in an exactly
  solved model and beyond},\ }\href
  {https://doi.org/https://doi.org/10.1016/j.aop.2005.10.005} {\bibfield
  {journal} {\bibinfo  {journal} {Annals of Physics}\ }\textbf {\bibinfo
  {volume} {321}},\ \bibinfo {pages} {2} (\bibinfo {year} {2006})}\BibitemShut
  {NoStop}%
\bibitem [{\citenamefont {He}\ \emph {et~al.}(2019)\citenamefont {He},
  \citenamefont {Ding}, \citenamefont {Zhou}, \citenamefont {Wang},\ and\
  \citenamefont {Gong}}]{He2019}%
  \BibitemOpen
  \bibfield  {author} {\bibinfo {author} {\bibfnamefont {A.-L.}\ \bibnamefont
  {He}}, \bibinfo {author} {\bibfnamefont {L.-R.}\ \bibnamefont {Ding}},
  \bibinfo {author} {\bibfnamefont {Y.}~\bibnamefont {Zhou}}, \bibinfo {author}
  {\bibfnamefont {Y.-F.}\ \bibnamefont {Wang}},\ and\ \bibinfo {author}
  {\bibfnamefont {C.-D.}\ \bibnamefont {Gong}},\ }\bibfield  {title} {\bibinfo
  {title} {Quasicrystalline chern insulators},\ }\href
  {https://doi.org/10.1103/PhysRevB.100.214109} {\bibfield  {journal} {\bibinfo
   {journal} {Phys. Rev. B}\ }\textbf {\bibinfo {volume} {100}},\ \bibinfo
  {pages} {214109} (\bibinfo {year} {2019})}\BibitemShut {NoStop}%
\bibitem [{\citenamefont {Chen}\ \emph {et~al.}(2023)\citenamefont {Chen},
  \citenamefont {Liu}, \citenamefont {Zou},\ and\ \citenamefont
  {Zhang}}]{Chen2023}%
  \BibitemOpen
  \bibfield  {author} {\bibinfo {author} {\bibfnamefont {Y.-k.}\ \bibnamefont
  {Chen}}, \bibinfo {author} {\bibfnamefont {Q.-h.}\ \bibnamefont {Liu}},
  \bibinfo {author} {\bibfnamefont {B.}~\bibnamefont {Zou}},\ and\ \bibinfo
  {author} {\bibfnamefont {Y.}~\bibnamefont {Zhang}},\ }\bibfield  {title}
  {\bibinfo {title} {Kitaev formula for periodic, quasicrystal, and fractal
  floquet topological insulators},\ }\href
  {https://doi.org/10.1103/PhysRevB.107.054109} {\bibfield  {journal} {\bibinfo
   {journal} {Phys. Rev. B}\ }\textbf {\bibinfo {volume} {107}},\ \bibinfo
  {pages} {054109} (\bibinfo {year} {2023})}\BibitemShut {NoStop}%
\bibitem [{\citenamefont {He}\ \emph {et~al.}(2013)\citenamefont {He},
  \citenamefont {Zhu}, \citenamefont {Wu}, \citenamefont {Liu}, \citenamefont
  {Liang},\ and\ \citenamefont {Kou}}]{He2013}%
  \BibitemOpen
  \bibfield  {author} {\bibinfo {author} {\bibfnamefont {J.}~\bibnamefont
  {He}}, \bibinfo {author} {\bibfnamefont {Y.-X.}\ \bibnamefont {Zhu}},
  \bibinfo {author} {\bibfnamefont {Y.-J.}\ \bibnamefont {Wu}}, \bibinfo
  {author} {\bibfnamefont {L.-F.}\ \bibnamefont {Liu}}, \bibinfo {author}
  {\bibfnamefont {Y.}~\bibnamefont {Liang}},\ and\ \bibinfo {author}
  {\bibfnamefont {S.-P.}\ \bibnamefont {Kou}},\ }\bibfield  {title} {\bibinfo
  {title} {Zero modes around vacancies in topological insulators and
  topological superconductors on the honeycomb lattice with particle-hole
  symmetry},\ }\href {https://doi.org/10.1103/PhysRevB.87.075126} {\bibfield
  {journal} {\bibinfo  {journal} {Phys. Rev. B}\ }\textbf {\bibinfo {volume}
  {87}},\ \bibinfo {pages} {075126} (\bibinfo {year} {2013})}\BibitemShut
  {NoStop}%
\bibitem [{\citenamefont {Bianco}\ and\ \citenamefont
  {Resta}(2011)}]{Resta2011}%
  \BibitemOpen
  \bibfield  {author} {\bibinfo {author} {\bibfnamefont {R.}~\bibnamefont
  {Bianco}}\ and\ \bibinfo {author} {\bibfnamefont {R.}~\bibnamefont {Resta}},\
  }\bibfield  {title} {\bibinfo {title} {Mapping topological order in
  coordinate space},\ }\href {https://doi.org/10.1103/PhysRevB.84.241106}
  {\bibfield  {journal} {\bibinfo  {journal} {Phys. Rev. B}\ }\textbf {\bibinfo
  {volume} {84}},\ \bibinfo {pages} {241106} (\bibinfo {year}
  {2011})}\BibitemShut {NoStop}%
\bibitem [{\citenamefont {Ceresoli}\ \emph {et~al.}(2006)\citenamefont
  {Ceresoli}, \citenamefont {Thonhauser}, \citenamefont {Vanderbilt},\ and\
  \citenamefont {Resta}}]{Resta2006}%
  \BibitemOpen
  \bibfield  {author} {\bibinfo {author} {\bibfnamefont {D.}~\bibnamefont
  {Ceresoli}}, \bibinfo {author} {\bibfnamefont {T.}~\bibnamefont
  {Thonhauser}}, \bibinfo {author} {\bibfnamefont {D.}~\bibnamefont
  {Vanderbilt}},\ and\ \bibinfo {author} {\bibfnamefont {R.}~\bibnamefont
  {Resta}},\ }\bibfield  {title} {\bibinfo {title} {Orbital magnetization in
  crystalline solids: Multi-band insulators, chern insulators, and metals},\
  }\href {https://doi.org/10.1103/PhysRevB.74.024408} {\bibfield  {journal}
  {\bibinfo  {journal} {Phys. Rev. B}\ }\textbf {\bibinfo {volume} {74}},\
  \bibinfo {pages} {024408} (\bibinfo {year} {2006})}\BibitemShut {NoStop}%
\end{thebibliography}%

\end{document}